\documentstyle[epsfig,graphicx,onecolumn]{mn}

\makeatletter
\def\@cite#1#2{(\if@tempswa #2 \fi #1)}
\makeatother

\def\etal{{et~al.}}
\def\eg{{e.g.}}
\def\ie{{i.e.}}
\def\pdif#1#2{\mathchoice{\partial{#1}\over\partial{#2}}
{\partial{#1}/\partial{#2}}{\partial{#1}/\partial{#2}}
{\partial{#1}/\partial{#2}}}
\newcommand{\bv}[1]{\mbox{$\bf #1$}}
\mathcode`\|="8000		%make | active in math mode
{\catcode`\|=\active \gdef|#1{_{\rm #1}}}

\title[Instabilities in two-fluid media]{Instabilities in two-fluid
magnetized media with inter-component drift}

\author[P.V. Tytarenko \etal{}]{P.V. Tytarenko$^{1,2}$\footnotemark, 
R.J.R. Williams$^{3}$\addtocounter{footnote}{-1}\footnotemark
and S.A.E.G. Falle$^{1}$\addtocounter{footnote}{-1}\footnotemark\\
$^{1}$ Department of Applied Mathematics, University of Leeds, Leeds LS2 9JT\\
$^{2}$ Astronomical Observatory of Kyiv Shevchenko University, 3 Observatorna
St., Kiev 04053, Ukraine\\
$^{3}$ Department of Physics and Astronomy, Cardiff University, PO Box 913,
Cardiff CF24 3YB}
\date{Received **INSERT**; in original form **INSERT**}
\pagerange{\pageref{firstpage}--\pageref{lastpage}}
\begin{document}
\label{firstpage}
\maketitle

\footnotetext{e-mail: pavlo@amsta.leeds.ac.uk,
robin.williams@astro.cf.ac.uk, sam@amsta.leeds.ac.uk}

\begin{abstract}
We analyse the stability of a magnetized medium consisting of a
neutral fluid and a fluid of charged particles, coupled to each other
through a drag force and exposed to differential body forces (for
example, as the result of radiation forces on one phase).  We consider
a uniform equilibrium and simple model input physics, but do not
arbitrarily restrict the relative orientations of the magnetic field,
slip velocity and wave vector of the disturbance.  We find several
instabilities and classify these in terms of wave resonances.  We
briefly apply our results to the structure of SiO maser regions
appearing in the winds from late-type stars.
\end{abstract}


\section{Introduction}

Multiphase flows are a widespread and important phenomenon in
astrophysics.  The difference between heating and cooling rates in
different components of astrophysical gases often leads to the
formation of a multicomponent medium, in which several phases with
widely separate temperatures coexist near to pressure equilibrium.
Effective multiphase behaviour can also result from differential
coupling of distinct particle species to local magnetic fields or
radiation driving forces.  To first order, these differential forces
will lead to drift velocities between the different components of the
fluid, limited by the effect of frictional terms.  However, this means
that there is local source of free energy in the flow.

Radiation pressure on dust, for example, plays a major role in many
models of the acceleration of winds from highly evolved, low-mass
stars \cite[\eg{}]{macgs92}.  The dust streams through the neutral gas
and transmits momentum to it through collisions.  The streaming of
dust through neutral gas has also received attention in many other
astrophysical contexts, including the evolution of dust bounded
H{\sc\,ii} regions \cite{cochran} and the radiation-driven implosion
of dense globules \cite{sanford}.  Radiation pressure on dust may
levitate interstellar clouds above the disc of the Milky Way
\cite{franco}: in such clouds dust particles will stream through the
neutral gas.  No magnetic field was included in any of these studies.
Hartquist \& Havnes~\shortcite{hartquist} identified conditions under
which dust grains are well-coupled to the magnetic field when the
grains are driven by radiation pressure.  In many circumstances the
dust, gas phase ions, and electrons may be treated as a single fluid.
Except for studies of the Wardle instability of shocks in dusty media
\cite{wardle90,stone97,macls97}, investigations of instabilities in
weakly ionized astrophysical media on lengthscales short compared to
the Jeans length driven by fluids streaming relative to one another
have been more limited.

There is a considerable literature on large-scale instabilities driven
by the self-gravity of multifluid media
\cite[\eg{}]{mous76,naka76,huba90,bz95,bals96,zw98,kn00,mams01}.
These papers differ in aspects such as the number of different flow
components assumed, the precise nature of the inter-species coupling
and the inclusion of processes such as the self-gravity of the flow
and large-scale gradients in flow properties.  Many of these papers
recover a large scale instability, first described by
Mouschovias~\shortcite{mous76} and Nakano~\shortcite{naka76}, in which
the diffusion of magnetic field out of a self-gravitating clump
reduces magnetic support, leading eventually to collapse.  In some,
rapidly growing, small lengthscale instabilities are found
\cite{huba90,kn00,mams01}, but, to date, rather restricted classes of
relative orientation of magnetic field, mean flow and wave vector have
been assumed.

We also note that there are many non-astronomical examples of
interspersed multiphase flows, such as clouds, fluidized beds and
microbial suspensions, which have been studied extensively.  For
example, Childress \& Spiegel~\shortcite{cs75} find buoyant
instabilities, similar to those we discuss here, in systems of finite
extent in both astrophysical and terrestrial contexts.

The past work on the Wardle instability and molecular cloud support
has treated inhomogeneous media, in which the streaming is induced,
for instance, by impulsive acceleration or by large scale variations
of the magnetic field.  While astrophysical flows are necessarily
inhomogeneous in the large, these variations can serve to obscure the
mechanisms of small-scale instability.  Given this wide variety of
physical mechanism and equilibrium structure, in this paper we outline
a general analysis for a simplified physical model, in which a charged
magnetized fluid streams through a neutral fluid as a result of
differential body forces.  This model might most directly be related
to flows with differential radiative forces on the fluids, but can be
applied more widely.  By assuming uniform initial conditions and
treating the modes which we find as distributions, we can study the
stability of short wavelength modes in general, without needing to
treat the specific global features which are important for longer
wavelengths.  Our analysis complements the previous work described
above, by giving stability criteria for wave-vectors of arbitrary
orientation and all initial angles between the body forces and
magnetic fields, albeit for rather simpler input physics.

In Section~\ref{s:basic}, we present the basic two-fluid equations and
derive the dispersion relation for linear waves.  In
Section~\ref{s:numeric}, we present numerical solutions of the
dispersion relation.  For small wavelengths, we find that `resonances'
(where distinct modes have similar phase velocities) are important in
understanding the stability properties, and discuss a graphical method
of locating these resonances in general geometries.  We then, in
Section~\ref{s:stability}, analyse the stability of the solutions of
the dispersion relation, proceeding from general analysis to specific
analytic stability criteria for short and long wavelengths.  These
criteria compare well with the numerical results in the previous
section, and confirm their generality.  In Section~\ref{s:wind} we
apply the long wavelength results to the properties of SiO maser spots
in late-type stars.  Finally, in Section~\ref{s:conclusion}, we
summarize our results.

\section{Basic equations and dispersion relation}

\label{s:basic}
In the present paper, we study the stability of two-fluid flows in
which one component is coupled to a magnetic field.  Differential
forces on the two fluids lead to inter-phase slip in the equilibrium
solution.  We attempt to characterise the general properties which
allow instabilities to feed off the slip energy.  While the system we
consider is simplified, it allows us to analyse the processes from
which instability results in some detail.

The equations we treat are those of continuity and momentum for
neutrals
\begin{eqnarray}
\pdif{\rho|n}{t} + \nabla.\left(\rho|n\bv{u}|n\right) &=& 0, \label{e:nmass}\\
\pdif{\bv{u}|n}{t} + \bv{u}|n.\nabla \bv{u}|n &=&
-{c|n^2\over \rho|n} \nabla \rho|n + \lambda\rho(\bv{u}-\bv{u}|n) - \bv{g},
\end{eqnarray}
and the ideal MHD equations for ions
\begin{eqnarray}
\pdif{\rho}{t} + \nabla.\left(\rho\bv{u}\right) &=& 0,\\
\pdif{\bv{u}}{t} + \bv{u}.\nabla \bv{u} &=&
-{c|i^2\over \rho} \nabla \rho
+{1\over\mu\rho}(\nabla\times\bv{B})\times\bv{B} + 
\lambda\rho|n(\bv{u}|n-\bv{u}) - \bv{g}|i,\\
\pdif{\bv{B}}{t} &=& \nabla\times(\bv{u}\times\bv{B})\label{e:dynamo},
\end{eqnarray}
where equation~(\ref{e:dynamo}) maintains the solenoidal condition
$\nabla.\bv{B} = 0$ so long as it is true initially, and the drag
terms correspond to Stokes' law with drag coefficient $\lambda$.  Here
$\rho|n$, $\bv{u}|n$, $c|n$ and $\bv{g}$ are, respectively, the mean
density, velocity, the effective isothermal sound speed and the net
acceleration for the neutrals; $\rho$, $\bv{u}$, $c|i$ and $\bv{g}|i$
are the equivalents for the ions; and $\lambda$ is a frictional
coupling constant.  The value of the vacuum permeability $\mu$
determines the magnetic field unit: conventional values include $1$,
$4\pi$ and $4\pi\times10^{-7}$.  These equations are similar to those
used in previous work \cite[\eg{}]{shu83,kn00}, except that in the
present paper we do not restrict the relative orientation of the
various vector fields (while, for the present, neglecting some of the
physical terms included by these earlier authors).

Two distinct pressure terms are used for the two distinct phases.
This might be taken as an assumption that the scattering between gas
particles of the same phase is far more rapid than that between
particles of differing phases.  The limits $c|i \to 0$ and $c|n\to 0$
are relevant in particular contexts, but have degenerate eigenmodes:
by assuming finite values of these parameters, the degeneracies are
lifted within the present analysis.

The forces on the ionized component are often dominated by the effects
of the overall curvature of the magnetic field, in which case one can
assume $\bv{u}_0.\bv{B} \simeq 0$ \cite{shu83,mous87}.  It is
consistent to study small-scale instabilities in the present of such
large scale gradients \cite[\eg{}]{huba90}.  However, in this case at
least part of the equilibrium force on one phase will not be
proportional to mass, as we have assumed above: the detailed
instability criteria will be somewhat different from those which we
derive below, but the general behaviour should be similar.  In what
follows, we study the equations for general orientations of the
various vector parameters, while bearing in mind the practical
importance of cases in which $\bv{u}_0.\bv{B}$ is small.

As in most previous papers, we have not included internal viscous
terms for the individual phases, although collisional processes will
in general lead to viscosity within the phases as well as inter-phase
drag.  This viscosity will lead to the stabilization of unstable wave
modes at short wavelengths.  We consider the effects of viscous terms
in the context of an astrophysical example in Section~\ref{s:visc},
and verify that, in that case, viscosity can be neglected at the
wavelengths which interest us.  We will present a detailed treatment
of the stability of flows including both inter-phase drag and internal
viscosity in a future paper.

We perturb about an equilibrium with a constant slip velocity
$\bv{u}_0 = \bv{u}-\bv{u}|n$ between the phases.  The slip velocity
satisfies the equilibrium condition
\begin{equation}
-\bv{g} + \lambda \rho \bv{u}_0 = -\bv{g}|i - \lambda \rho|n \bv{u}_0 = 0.
\end{equation}

Linearizing about this equilibrium with $\bv{u}|n = \bv{v}|n$,
$\bv{u} = \bv{u}_0+\bv{v}$, $\rho = \rho_0 + \theta$, $\rho|n =
\rho|{n,0} + \theta|n$ and $\bv{B} = \bv{B}_0 + \bv{\beta}$, we find
\begin{eqnarray}
\dot\theta|n + \rho|{n,0}\nabla.\bv{v}|n &=& 0,\\
\dot{\bv{v}}|n &=& -{c|n^2\over\rho|{n,0}}\nabla\theta|n
	+ \lambda\rho_0(\bv{v}-\bv{v}|n) + \lambda\theta\bv{u}_0, \\
\dot\theta + \bv{u}_0.\nabla\theta+\rho_0\nabla.\bv{v} &=& 0,\\
\dot{\bv{v}} + \bv{u}_0.\nabla\bv{v} &=&
	-{c|i^2\over\rho|{0}}\nabla\theta
	-\lambda\theta|n\bv{u}_0 + \lambda\rho|{n,0}(\bv{v}|n-\bv{v})
	+{1\over\mu\rho_0}(\nabla\times\bv{\beta})\times\bv{B}_0,\\
\dot{\bv\beta} &=& -\bv{u}_0.\nabla\beta + (\bv{B}_0.\nabla) \bv{v}
- \bv{B}_0(\nabla.\bv{v}).
\end{eqnarray}
In what follows we will suppress indices 0 on $\rho_0$ and
$\rho|{n,0}$.  Looking for solutions of form $\exp
i(\bv{k}.\bv{x}-\omega t)$, we find
\begin{eqnarray}
-\omega\theta|n &=& - \rho|n\bv{k}.\bv{v}|n,
\label{e:lintn}\\
(-\omega-i\lambda\rho)\bv{v}|n &=& -{c|n^2\over\rho|n}\bv{k}\theta|n
	-i\lambda(\rho\bv{v}+\bv{u}_0\theta),
\label{e:linvn}\\
(\bv{u}_0.\bv{k}-\omega)\theta &=& - \rho\bv{k}.\bv{v},
\label{e:linth}\\
(\bv{u}_0.\bv{k}-\omega-i\lambda\rho|n)\bv{v} &=&
	-{c|i^2\over\rho}\bv{k}\theta
	-i\lambda(\rho|n\bv{v}|n-\bv{u}_0\theta|n)
	+{1\over\mu\rho}\left[
		(\bv{B}_0.\bv{k})\bv{\beta}-\bv{k}(\bv{B}_0.\bv{\beta})
	\right],
\label{e:linv}\\
(\bv{u}_0.\bv{k}-\omega\bv)\bv{\beta} &=& 
(\bv{B}_0.\bv{k}) \bv{v} - \bv{B}_0(\bv{k}.\bv{v}).
\label{e:linbt}
\end{eqnarray}
These equations may now be manipulated to give a dispersion relation
in the form ${\cal D}(\omega,\bv{k}) = 0$, either directly or by
noticing that they take the form of an eigenequation for eigenvalue
$\omega$, with eigenvector $\bv{U}^{\rm T} = (\theta|n,\bv{v}|n^{\rm
T},\theta,\bv{v}^{\rm T},\bv{\beta}^{\rm T})$.

Some factors of the dispersion relation can easily be found from
equations~(\ref{e:lintn}--\ref{e:linbt}).  Taking the scalar product
of $\bv{k}$ with (\ref{e:linbt}), we find
\begin{equation}
(\omega-\bv{u}_0.\bv{k})\bv{\beta}.\bv{k} = 0,
\end{equation}
corresponding to an eigenvalue $\omega = \bv{u}_0.\bv{k}$ with
eigenvector components $\bv{\beta}$ parallel to $\bv{k}$ and $\theta|n
= \bv{v}|n = \theta = \bv{v} = 0$.  This trivial eigenvector results
from the requirement that $\nabla.\bv{B}=0$ be maintained, and has an
amplitude identically zero.

A further factor can be found, for which $\theta|n = \theta = 0$.
Using equations~(\ref{e:lintn}--\ref{e:linbt}), we find that
\begin{equation}
\left[(\omega-\bv{u}_0.\bv{k})^2 - {(\bv{B}_0.\bv{k})^2\over\mu\rho}\right]
\omega+
i\lambda\left\{\rho
\left[(\omega-\bv{u}_0.\bv{k})^2 - {(\bv{B}_0.\bv{k})^2\over\mu\rho}\right]
+ \rho|n\omega(\omega-\bv{u}_0.\bv{k})\right\} = 0,
\label{e:gpa}
\end{equation}
with the components of the eigenvector having $\bv{v}$ parallel to
$\bv{B}_0\times\bv{k}$,
\begin{eqnarray}
\bv{v}|n &=& {i\lambda\rho\over\omega+i\lambda\rho}\bv{v},\\
\bv{\beta} &=& -{\bv{B}_0.\bv{k}\over\omega-\bv{u}_0.\bv{k}}\bv{v}.
\end{eqnarray}
In these modes, two Alfv\'en waves in the ionized gas are coupled to a
shear mode in the neutral gas.  We will refer to these as group A
modes.

The remaining 7 roots of the dispersion relation, which we will refer
to as group B modes, are coupled by the remaining 2 components of the
drag force.  To simplify the form of the dispersion relation somewhat,
we assume (without loss of generality) that the wave vector $\bv{k}$
is parallel to $\hat{\bv{x}}$, and introduce scaled variables and
parameters as follows
\begin{equation}
\Omega=\frac{\omega}{k}, \;
\Omega|i=\frac{\lambda \rho}{k}, \;
\Omega|n=\frac{\lambda \rho|n}{k}, 
\end{equation}
and a vector, $\bv{v}|A$, with magnitude equal to the ionic Alfv\'en
speed and direction parallel to $\bv{B}_0$,
\begin{equation}
\bv{v}|A={\bv{B}_0\over \sqrt{\mu\rho}}.
\end{equation}
Note that each of these variables has dimensions of velocity.  This
proves useful when interpreting the results of our analysis.

In terms of these variables, the dispersion relation for the group A
modes, equation~(\ref{e:gpa}), takes the form
\begin{equation}
\left[ (\Omega- u_x)^2 - v_{{\rm A}x}^2 \right] \Omega
+i\Omega|i\left[ (\Omega- u_x)^2 - v_{{\rm A}x}^2 \right]
+i\Omega|n\Omega(\Omega-u_x)=0.\label{e:gpan}
\end{equation}

The dispersion relation for the group B modes, obtained from equations
(\ref{e:lintn}--\ref{e:linbt}), has the form
\begin{eqnarray}
(\Omega-u_x) \left[ (\Omega-u_x)(\Omega+
i\Omega|i)+i\Omega \Omega|n \right] \left\{ (\Omega^2-c|n^2+
i\Omega \Omega|i) \left[ (\Omega-u_x)^2-
(v^2_{\rm A}+c|i^2) \right]+ i\Omega|n(\Omega-u_x)(\Omega^2-c|n^2)\right\}
		\nonumber\\
-\Omega|i\Omega|n c|n^2 v_{{\rm A}x} \left[
\Omega v_{{\rm A}x}-\bv{v}|A.\bv{u} \right]+
c|i^2v_{{\rm A}x}^2 (\Omega+i\Omega|i)
(\Omega^2-c|n^2+i\Omega \Omega|i)&=&0.\label{e:gpb}
\end{eqnarray}
It is useful to write this latter dispersion relation in the form
\begin{equation}
P_0(\Omega)+i\left[ \Omega|i P_1(\Omega)+\Omega|n 
P_2(\Omega) \right]-
\Omega|i^2 P_{11}(\Omega)-
\Omega|i \Omega|n P_{12}(\Omega)
-\Omega|n^2 P_{22}(\Omega)=0,\label{e:pdisp}
\end{equation}
where
\begin{eqnarray}
P_0(\Omega)&=&\Omega(\Omega^2-c|n^2) \left[ (\Omega-
u_x)^2-v|f^2 \right] \left[ 
(\Omega-u_x)^2 -v|s^2\right],\label{e:p0}\\
P_1(\Omega)&=&(2\Omega^2-c|n^2) \left[ (\Omega-u_x)^2-v|f^2 \right] 
\left[ (\Omega-u_x)^2-
v|s^2\right],\\
P_2(\Omega)&=&\Omega(\Omega^2-c|n^2)(\Omega-
u_x) \left[ 2(\Omega-u_x)^2- (v^2_{\rm A}+c|i^2) \right],\\
P_{11}(\Omega)&=&\Omega \left[ (\Omega-u_x)^2-v|f^2
\right] \left[ (\Omega-u_x)^2-v|s^2 \right]
= \Omega\left[(\Omega-u_x)^4-(v|A^2+c|i^2)(\Omega-u_x)^2
		+c|i^2v_{{\rm A}x}^2\right],\label{e:p11}
\\
P_{12}(\Omega)&=& 2\Omega^2(\Omega-u_x)^3-(v|A^2+c|i^2)\Omega^2(\Omega-u_x) 
-c|n^2\left[(\Omega-u_x)^3
-\Omega v_{{\rm A}x}^2+\bv{v}|A.\bv{u}v_{{\rm A}x} \right],
\label{e:p12}\\
P_{22}(\Omega)&=&\Omega(\Omega-u_x)^2(\Omega^2-c|n^2),\label{e:p22}
\end{eqnarray}
and
\begin{equation}
v|{f,s}^2=\frac{1}{2} \left\{v|A^2+c|i^2 \pm \sqrt{(v|A^2+c|i^2)^2-4c|i^2
v_{{\rm A}x}^2} \right\}
\end{equation}
are the fast and slow ionic magnetosound wave speeds.

\section{Numerical results}

\label{s:numeric}
\begin{figure*}
\begin{centering}
\begin{tabular}{ll}
(a) & (b) \\
\epsfysize=8cm\rotatebox{270}{\epsffile{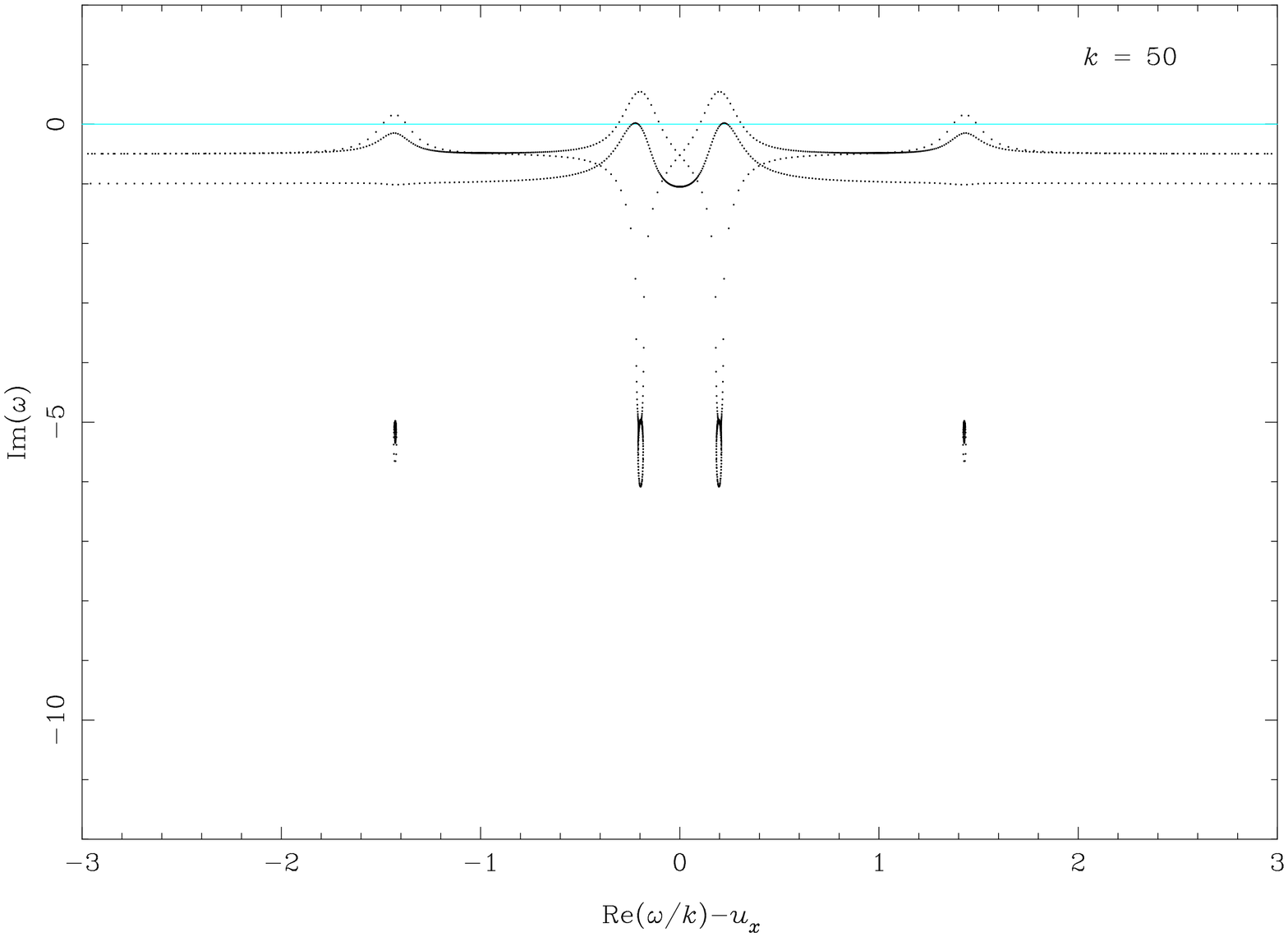}} &
\epsfysize=8cm\rotatebox{270}{\epsffile{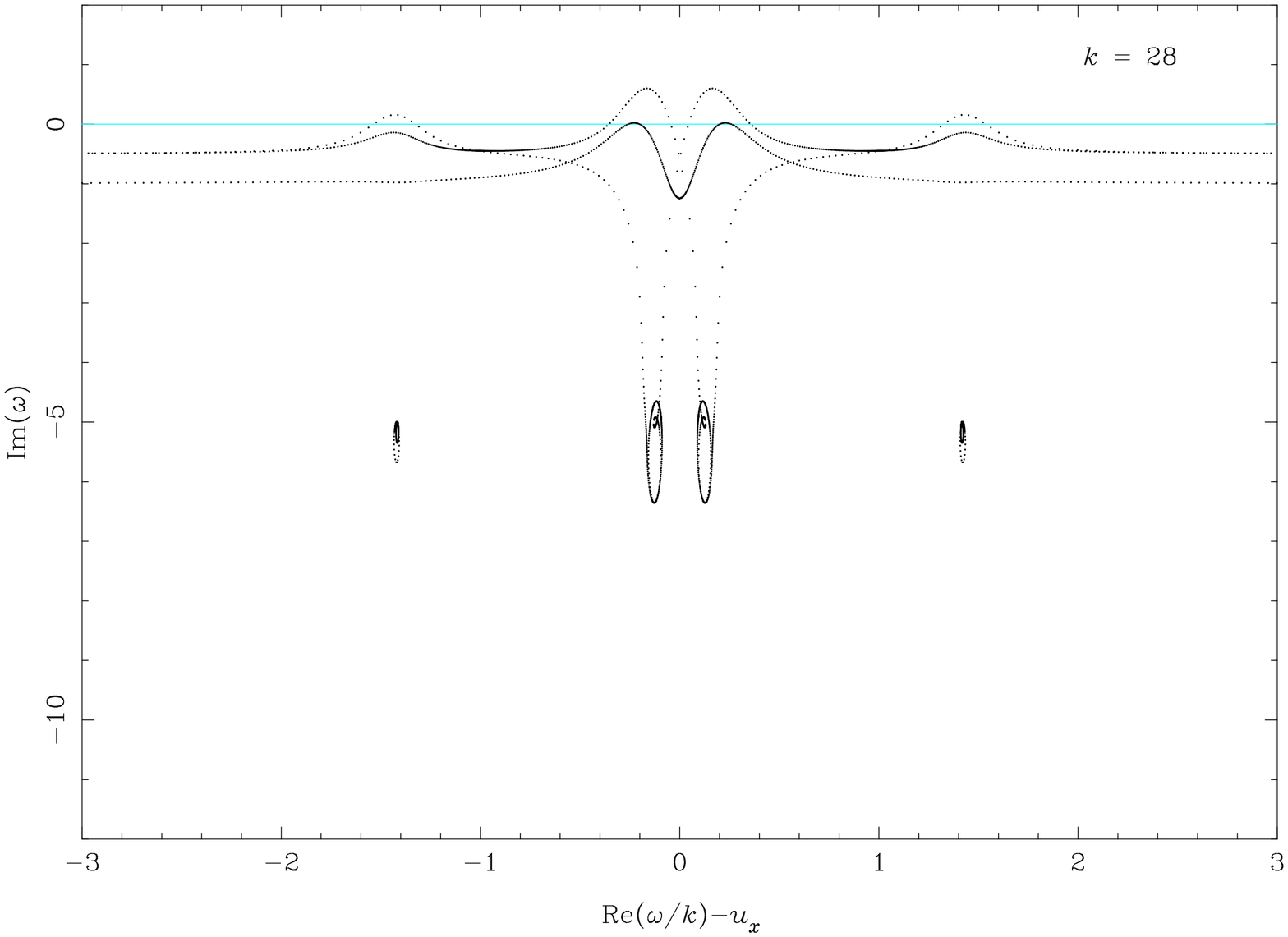}} \\
(c) & (d) \\
\epsfysize=8cm\rotatebox{270}{\epsffile{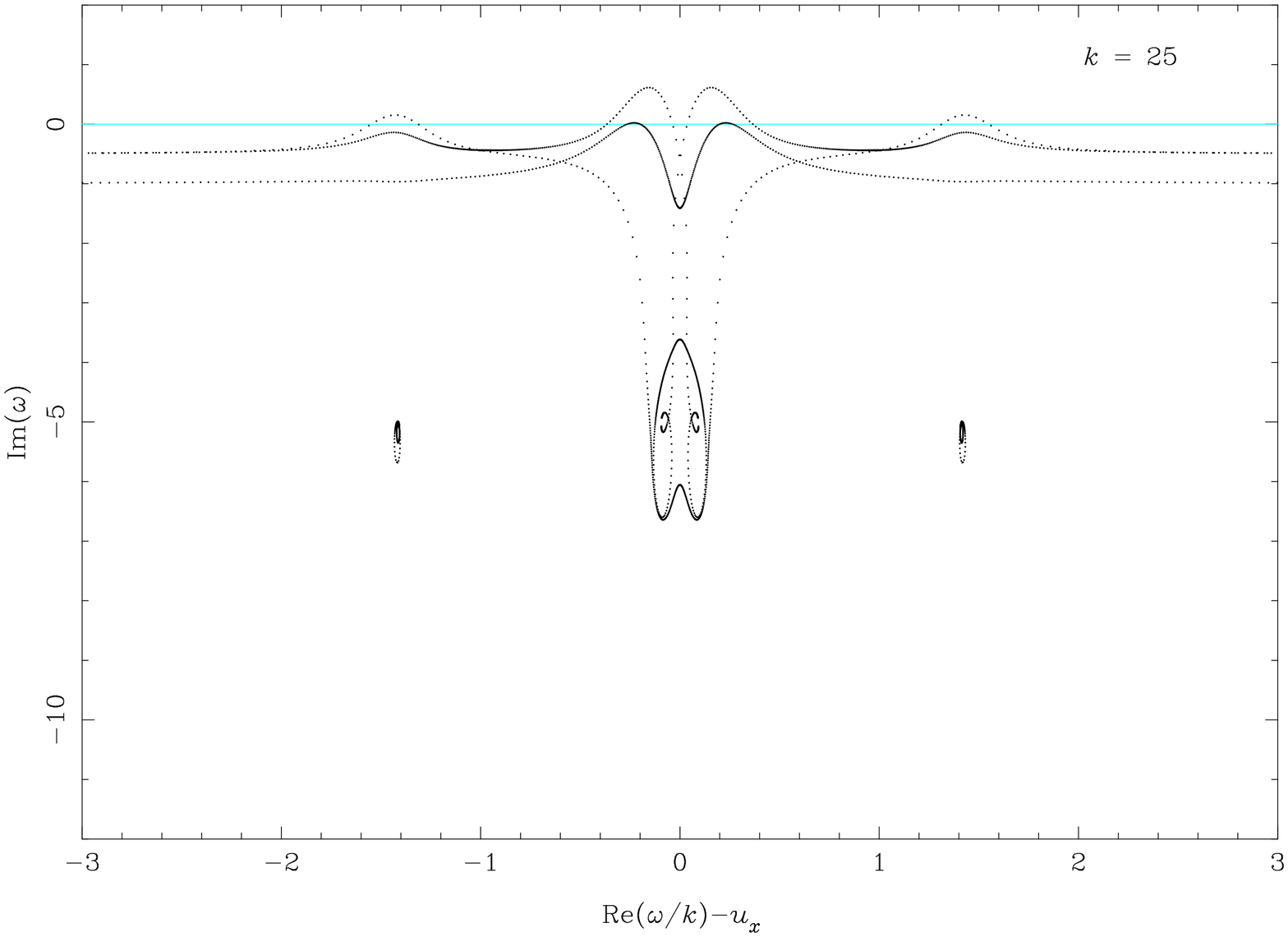}} &
\epsfysize=8cm\rotatebox{270}{\epsffile{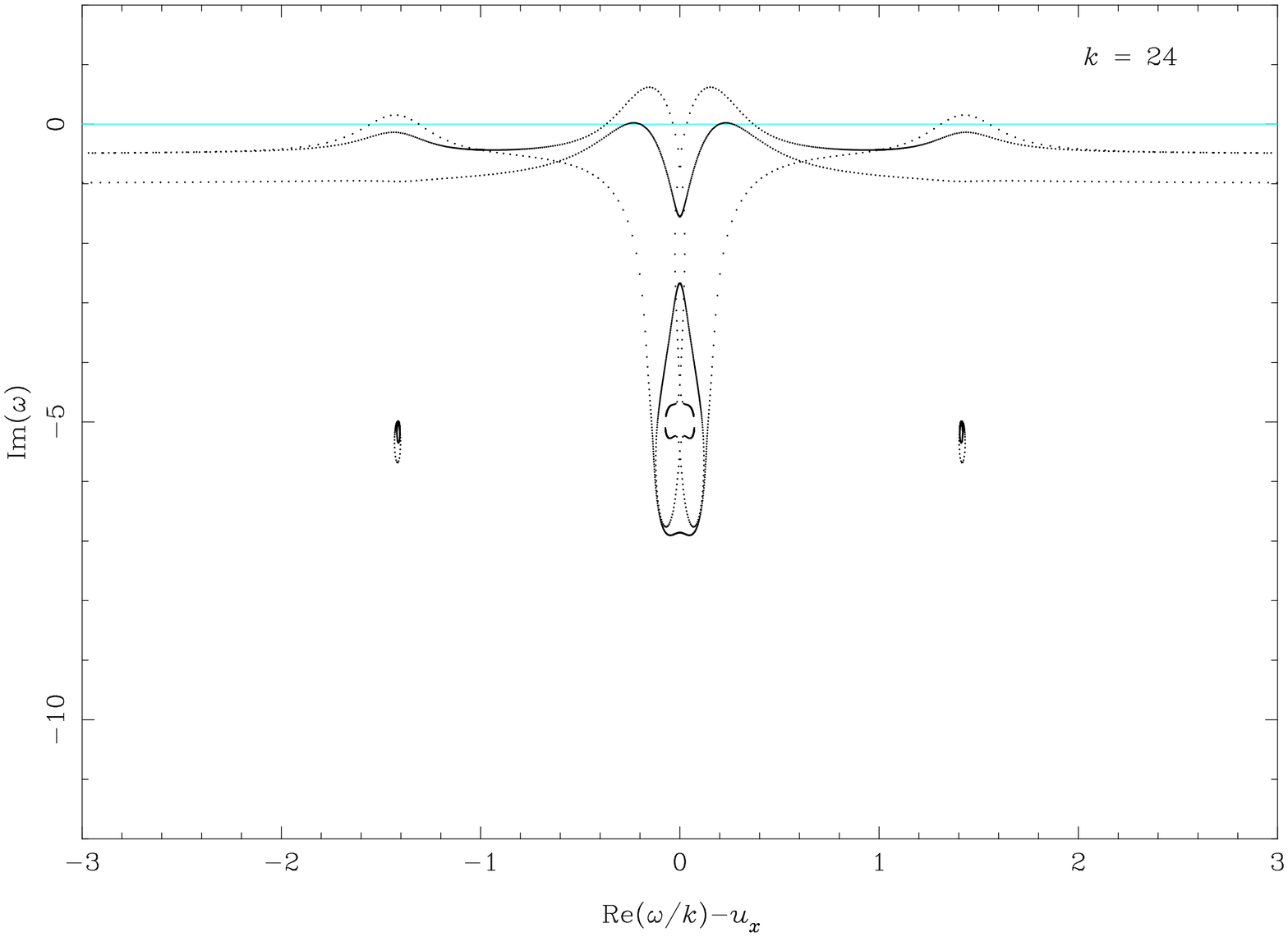}} \\
(e) & (f) \\
\epsfysize=8cm\rotatebox{270}{\epsffile{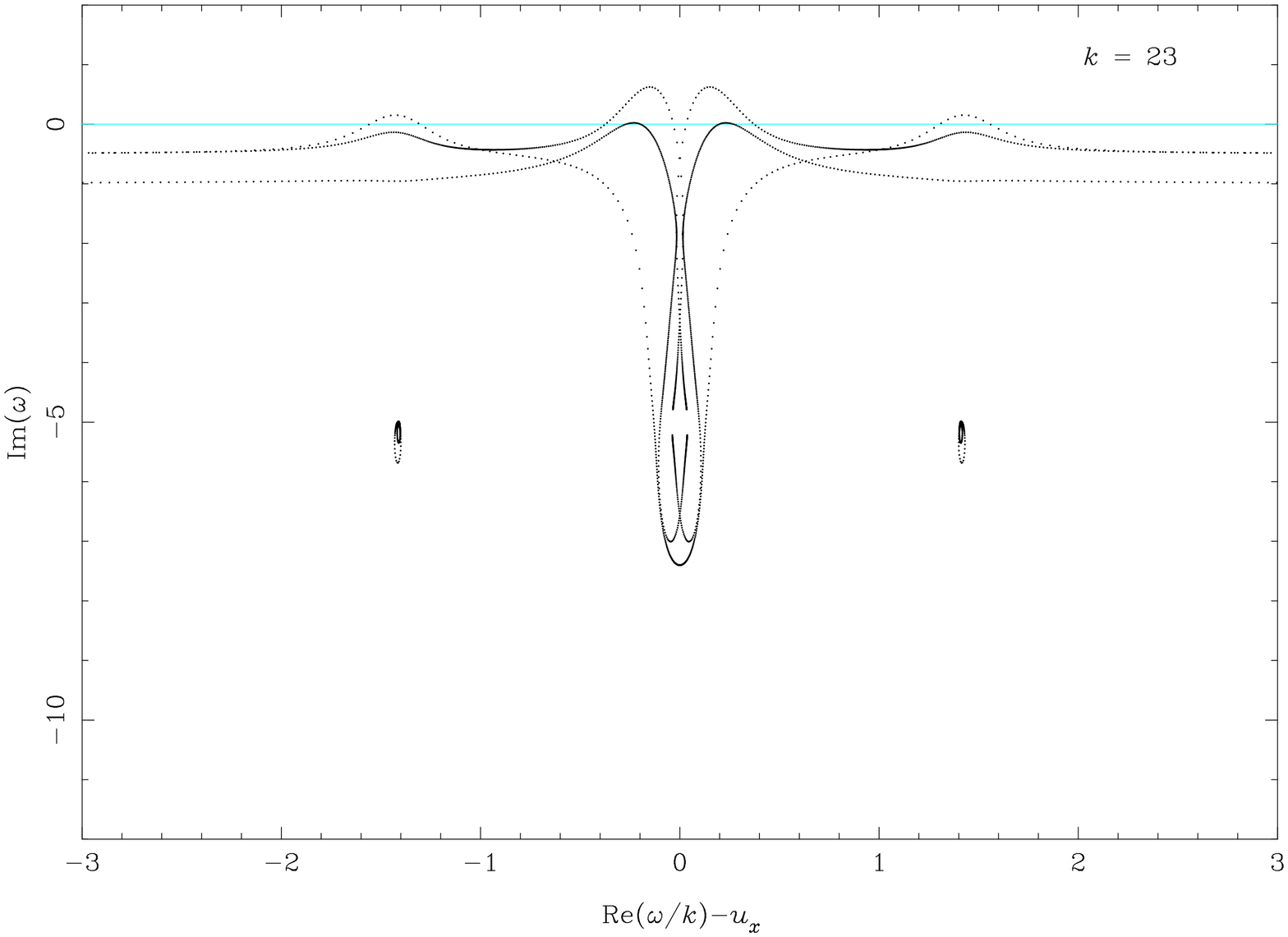}} &
\epsfysize=8cm\rotatebox{270}{\epsffile{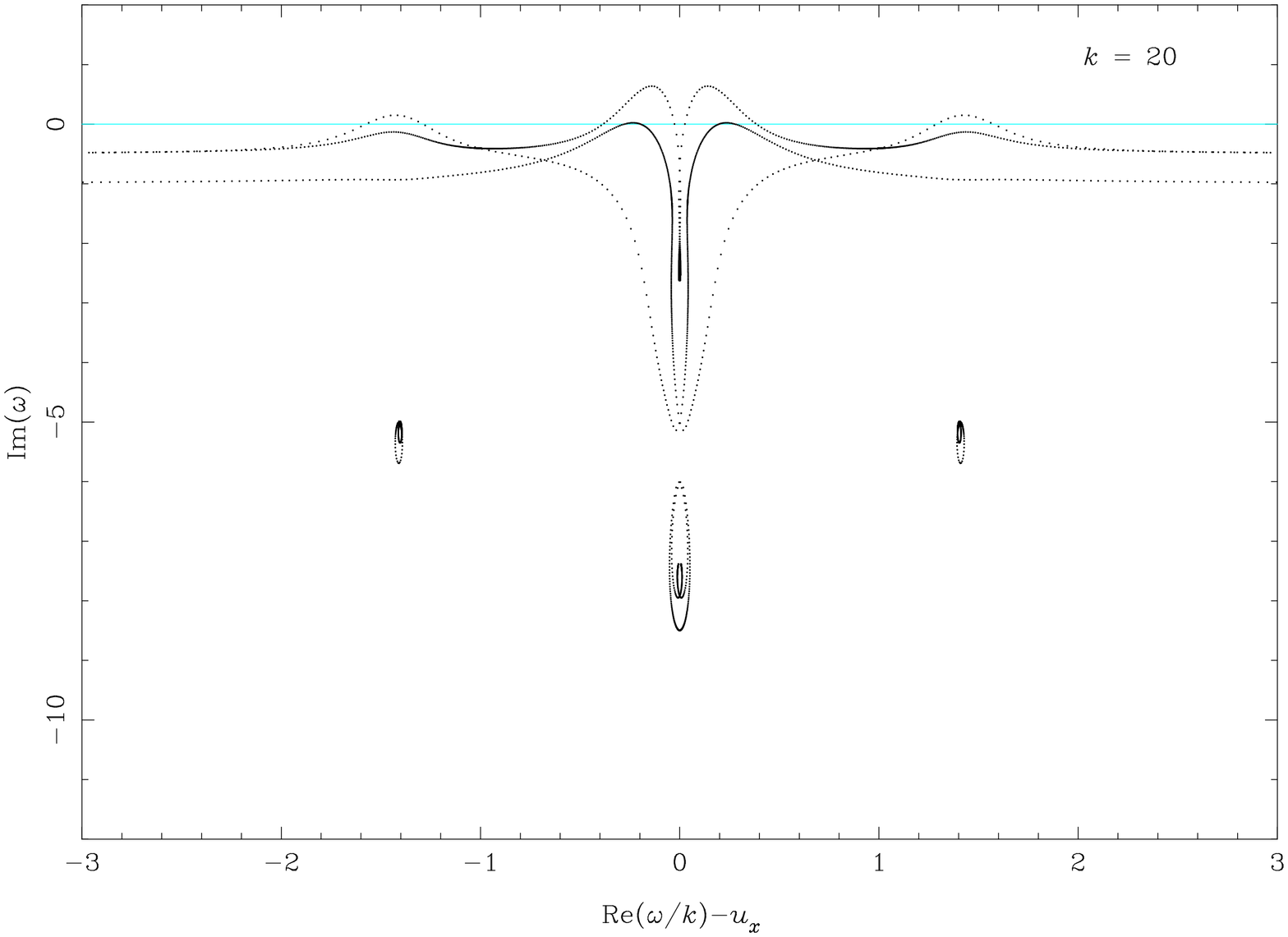}} 
\end{tabular}
\end{centering}
\caption[]{Variation of eigenvalues of group B with $u_x$, for various
$k$.  In this example, $\bv{B} = (1,0,1)\sqrt{\mu}$, $\lambda = 1$,
$\rho = 1$, $\rho|n = 10$, $c|i^2 = 0.1$, $c|n^2 = 1$, $\bv{k} = (1,0,0)k$ 
and $\bv{u}=(1,0,-1)u_x$.}
\label{f:ex1}
\end{figure*}

\begin{figure*}
\begin{centering}
\begin{tabular}{ll}
(g) & (h) \\
\epsfysize=8cm\rotatebox{270}{\epsffile{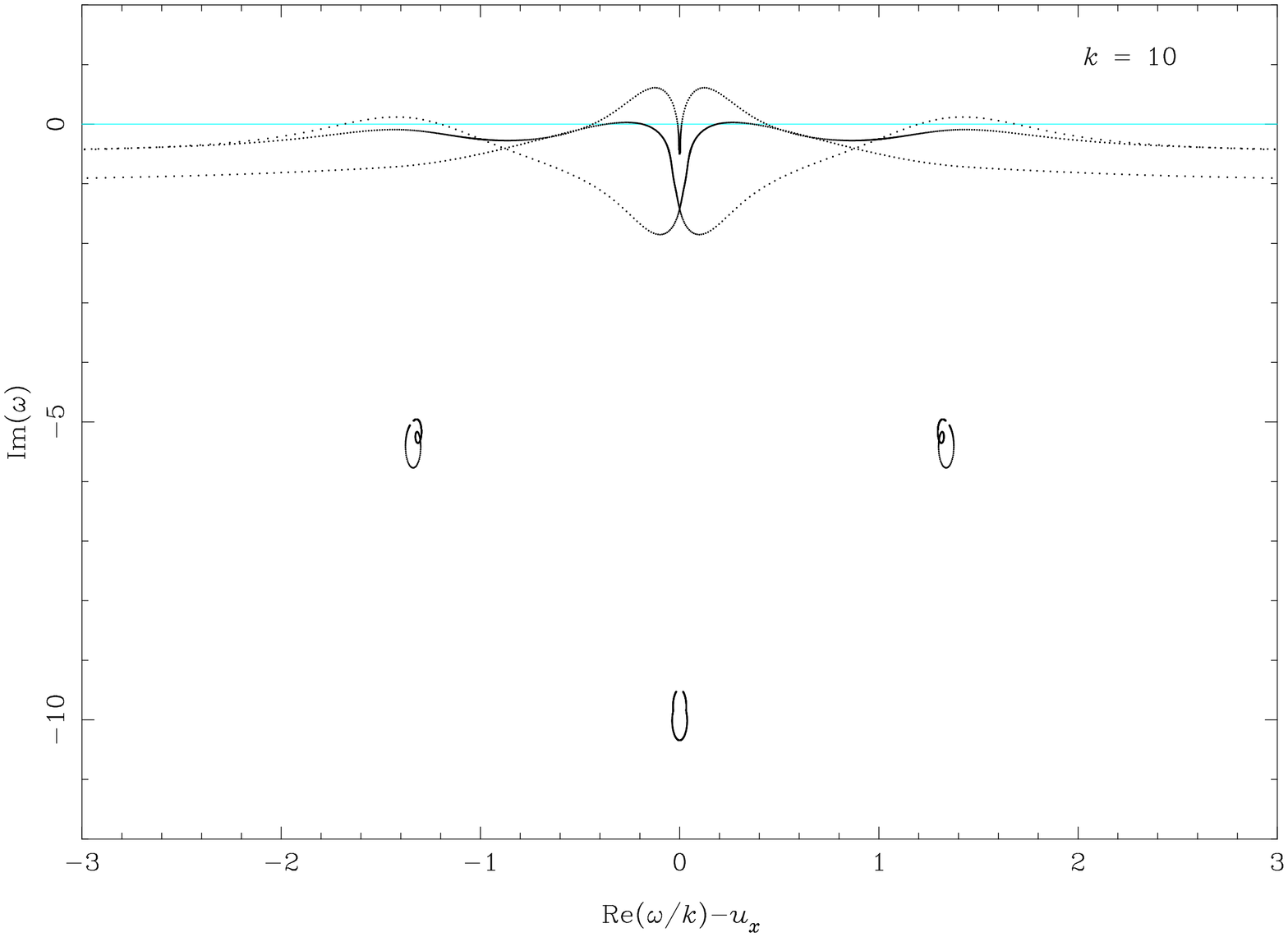}} &
\epsfysize=8cm\rotatebox{270}{\epsffile{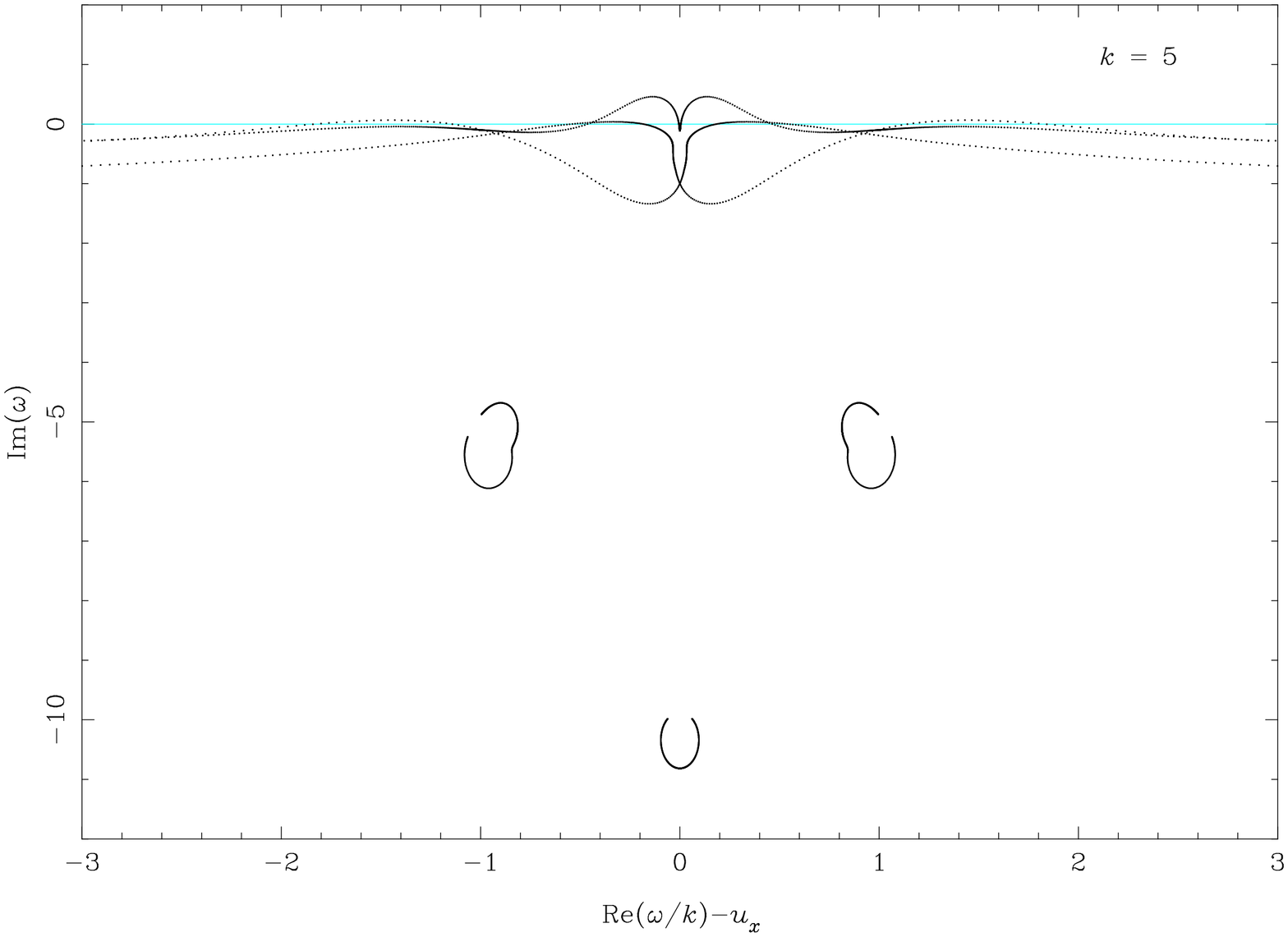}} \\
(i) & (j) \\
\epsfysize=8cm\rotatebox{270}{\epsffile{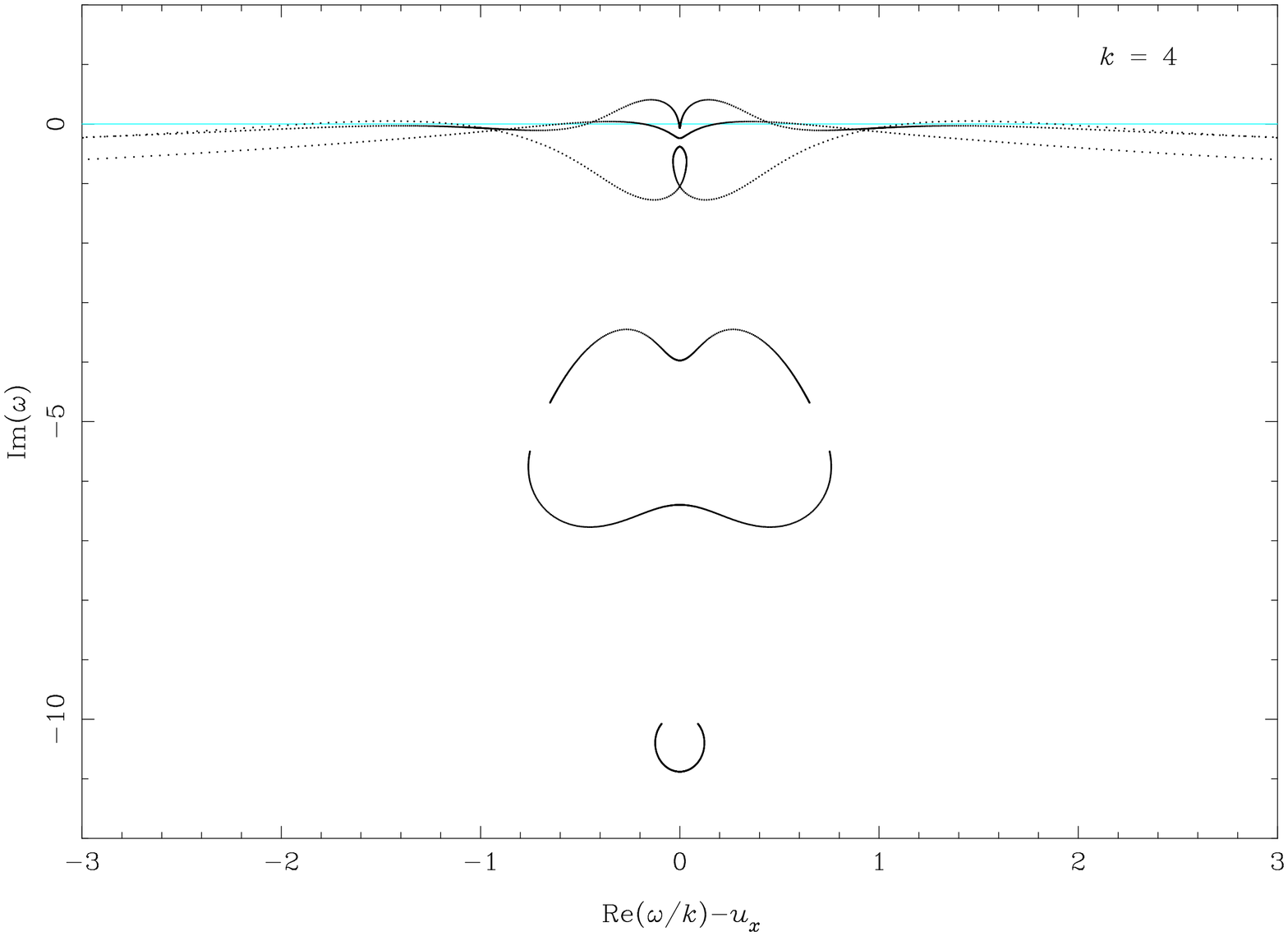}} &
\epsfysize=8cm\rotatebox{270}{\epsffile{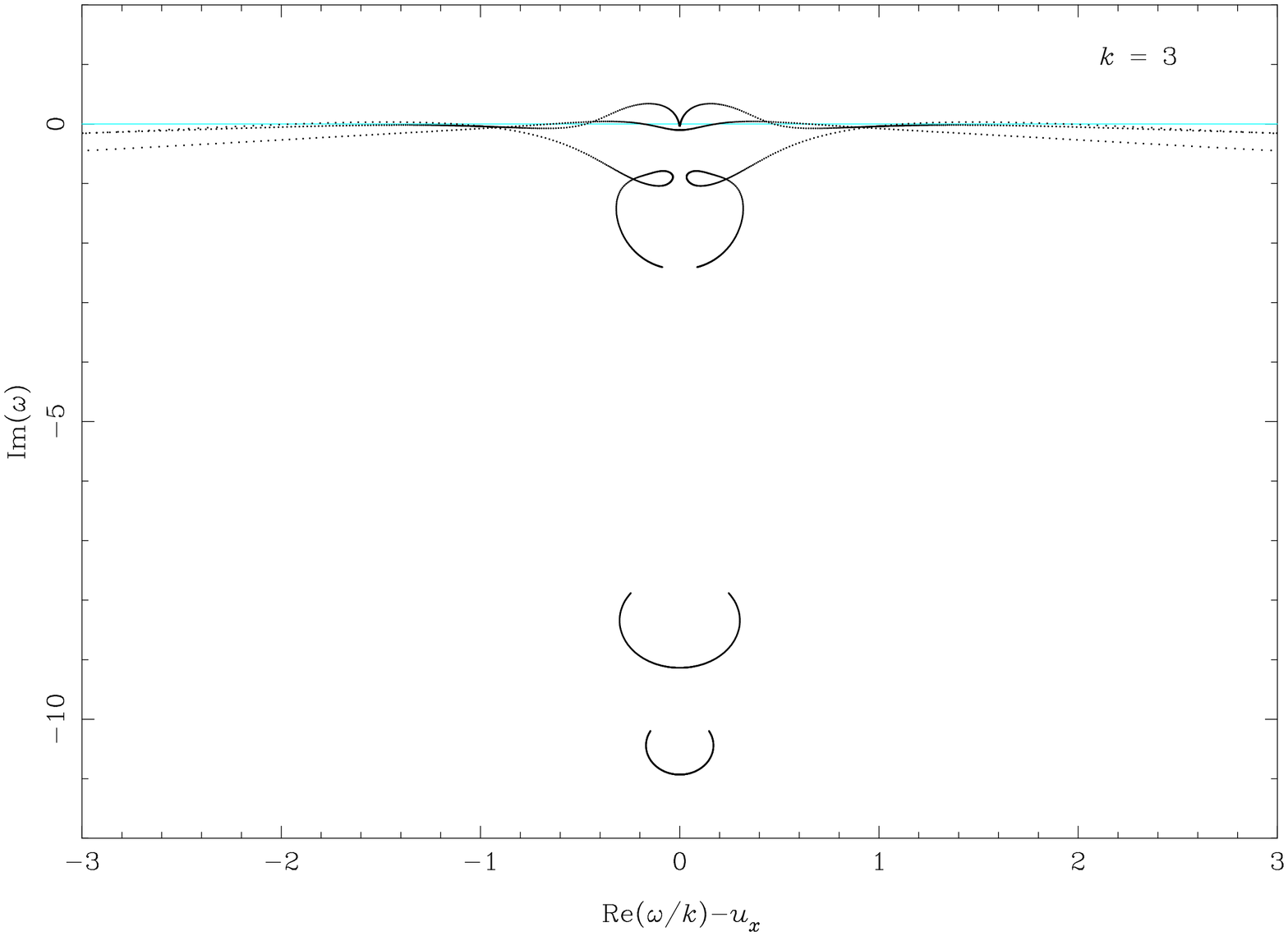}} \\
(k) & (l) \\
\epsfysize=8cm\rotatebox{270}{\epsffile{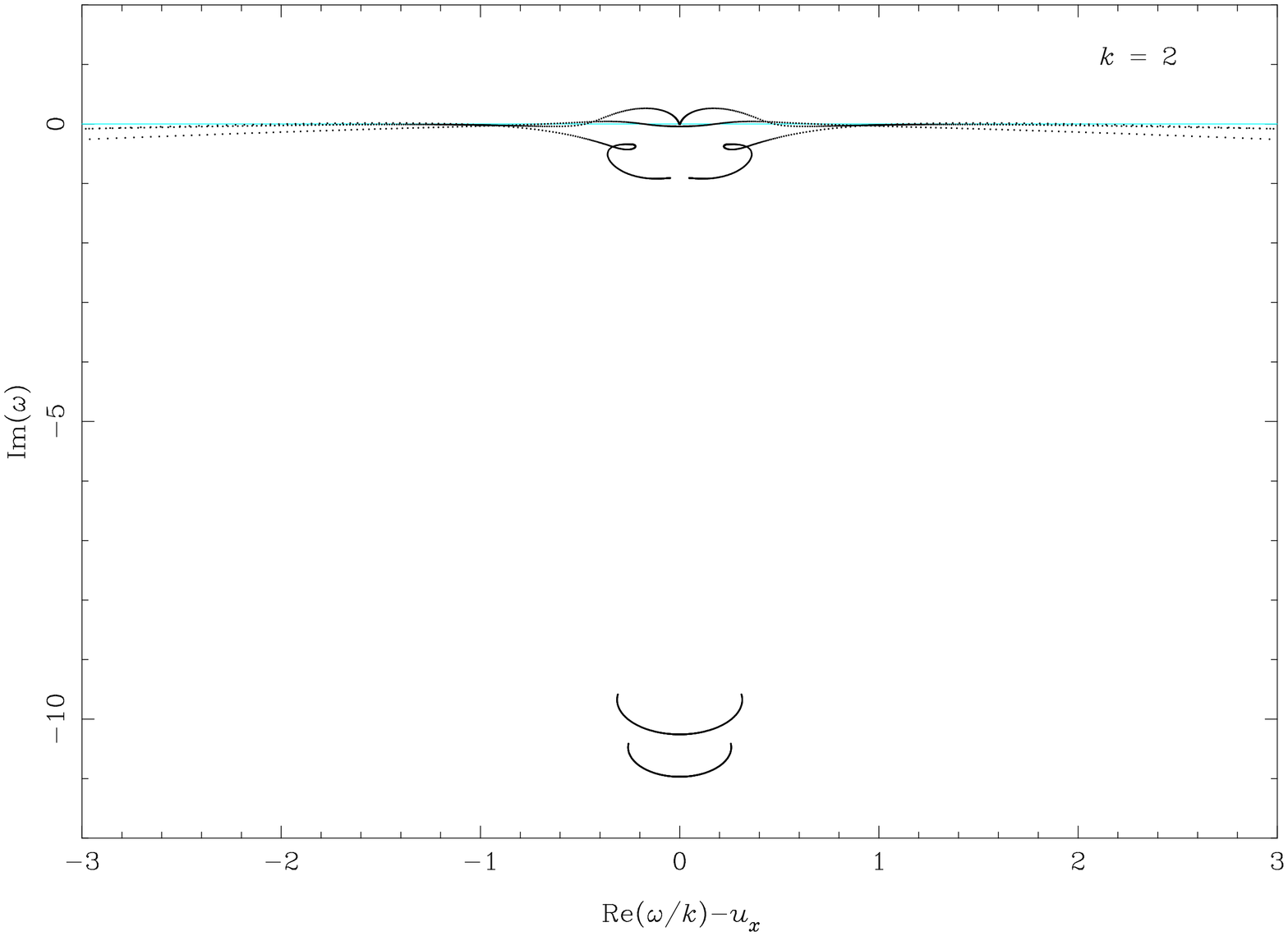}} &
\epsfysize=8cm\rotatebox{270}{\epsffile{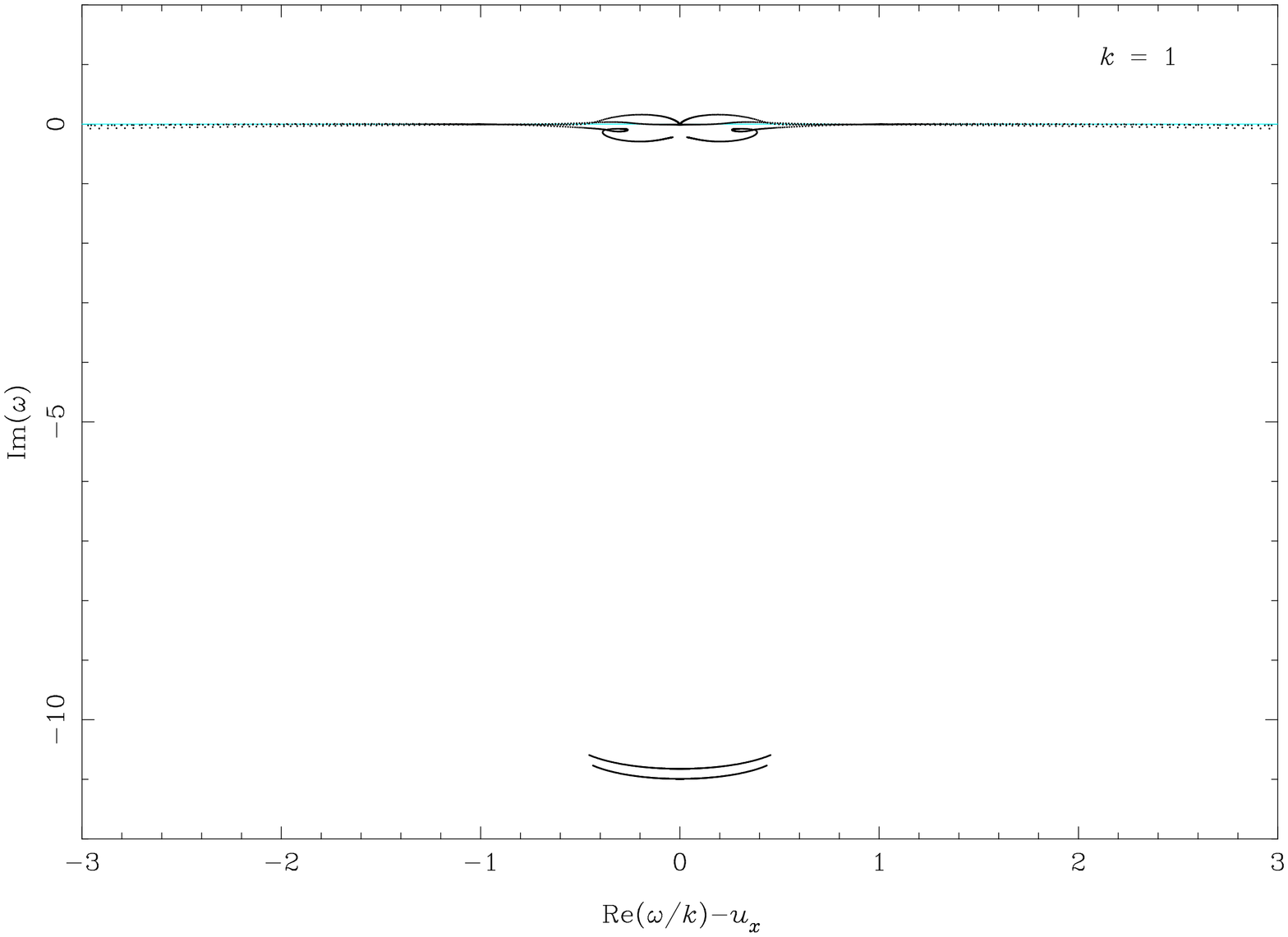}} 
\end{tabular}
\end{centering}
\contcaption{}
\end{figure*}

\begin{figure*}
\begin{centering}
\begin{tabular}{ll}
\multicolumn{2}{c}{
\begin{tabular}{l}
(a)\\\epsfysize=7cm\rotatebox{270}{\epsffile{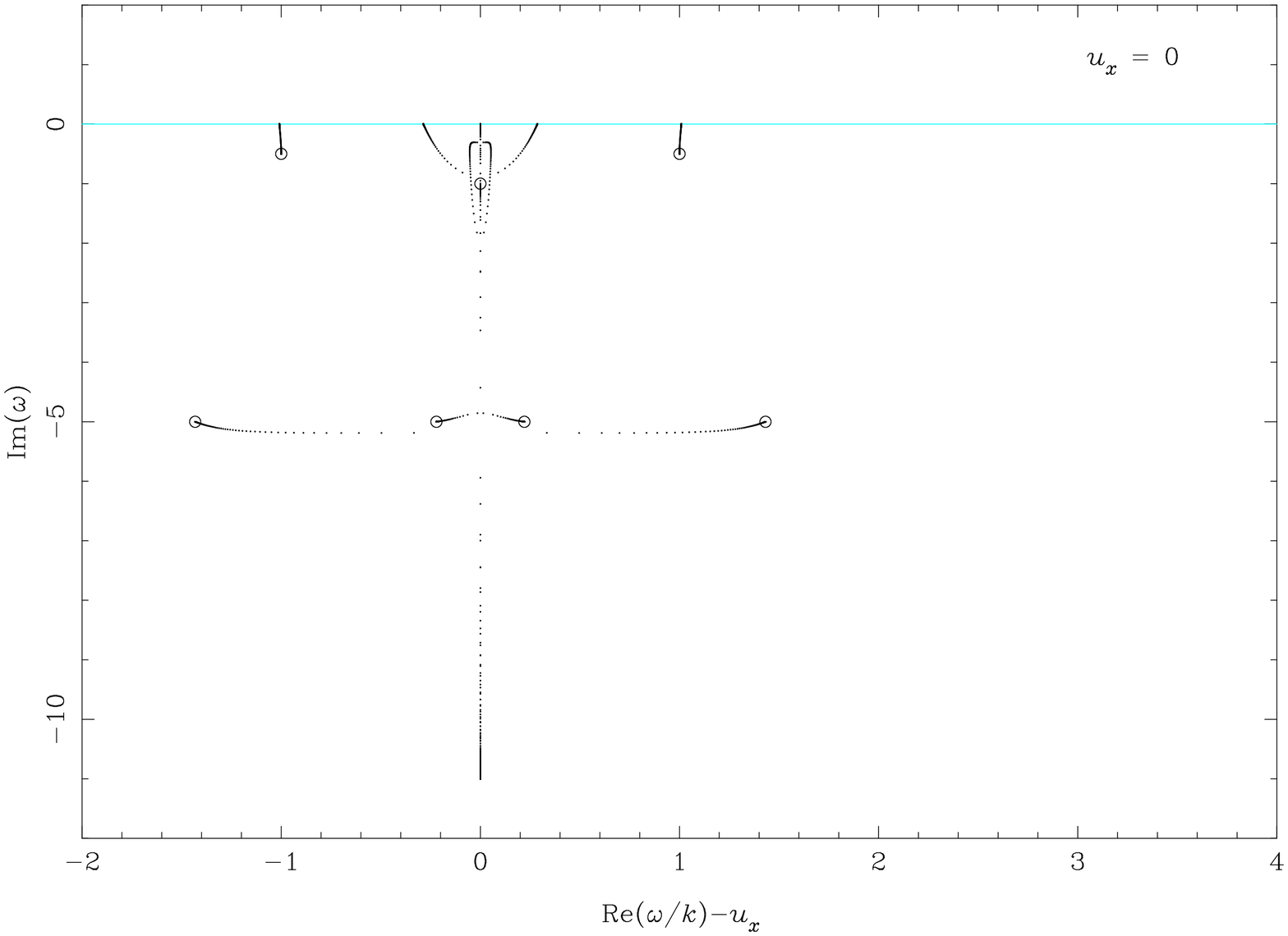}}
\end{tabular}}\\
(b) & (c) \\
\epsfysize=7cm\rotatebox{270}{\epsffile{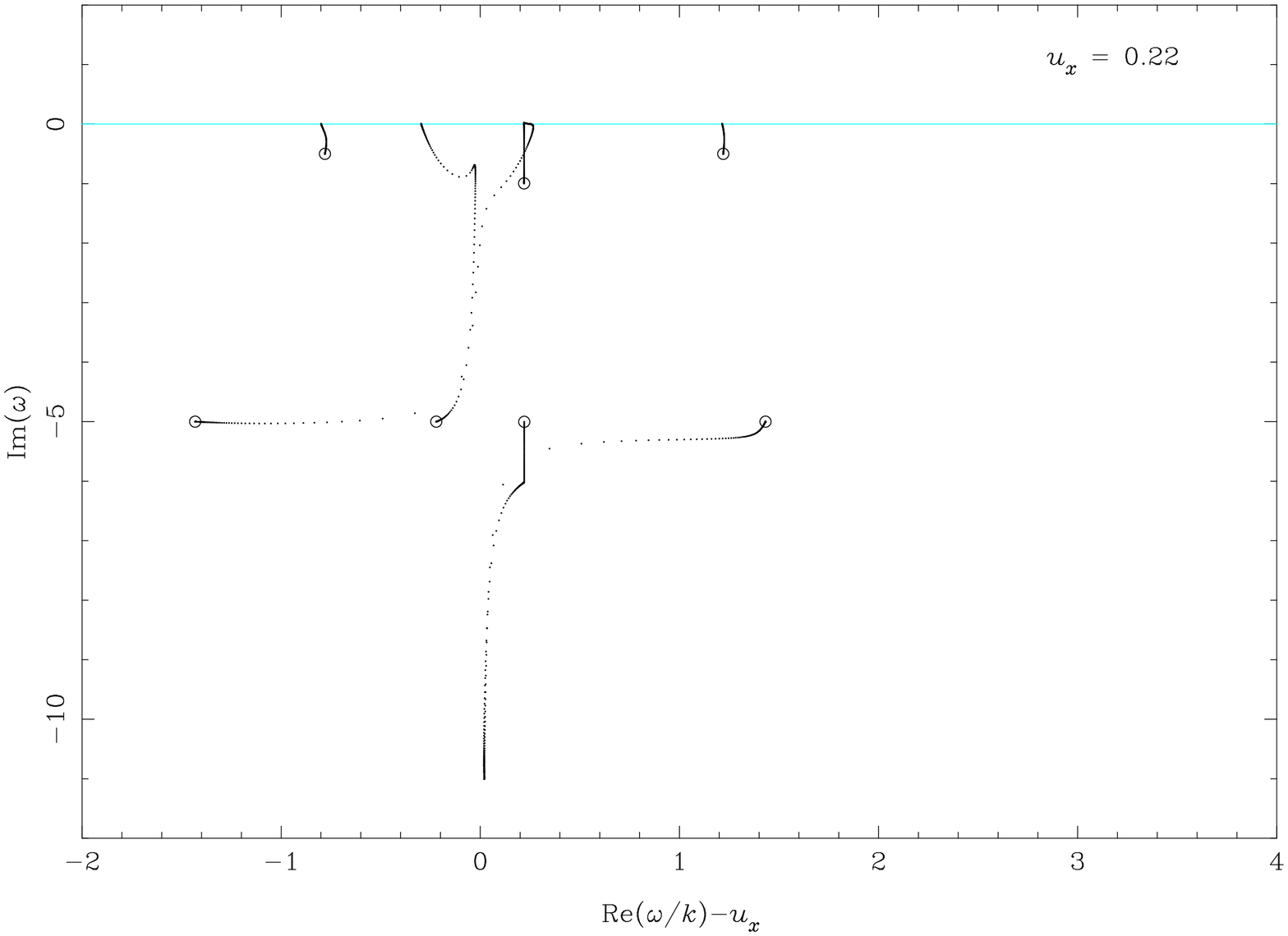}} &
\epsfysize=7cm\rotatebox{270}{\epsffile{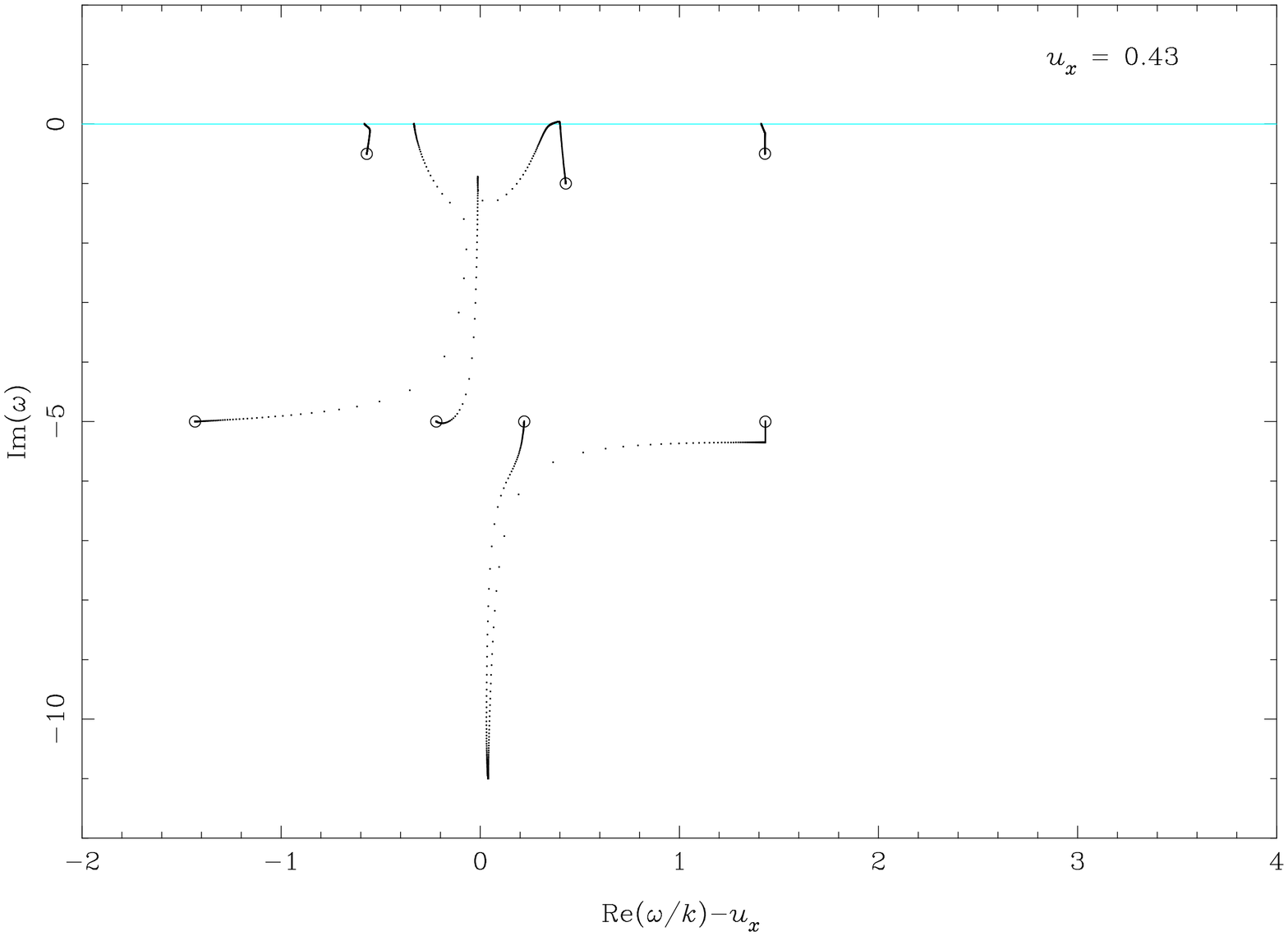}} \\
(d) & (e) \\
\epsfysize=7cm\rotatebox{270}{\epsffile{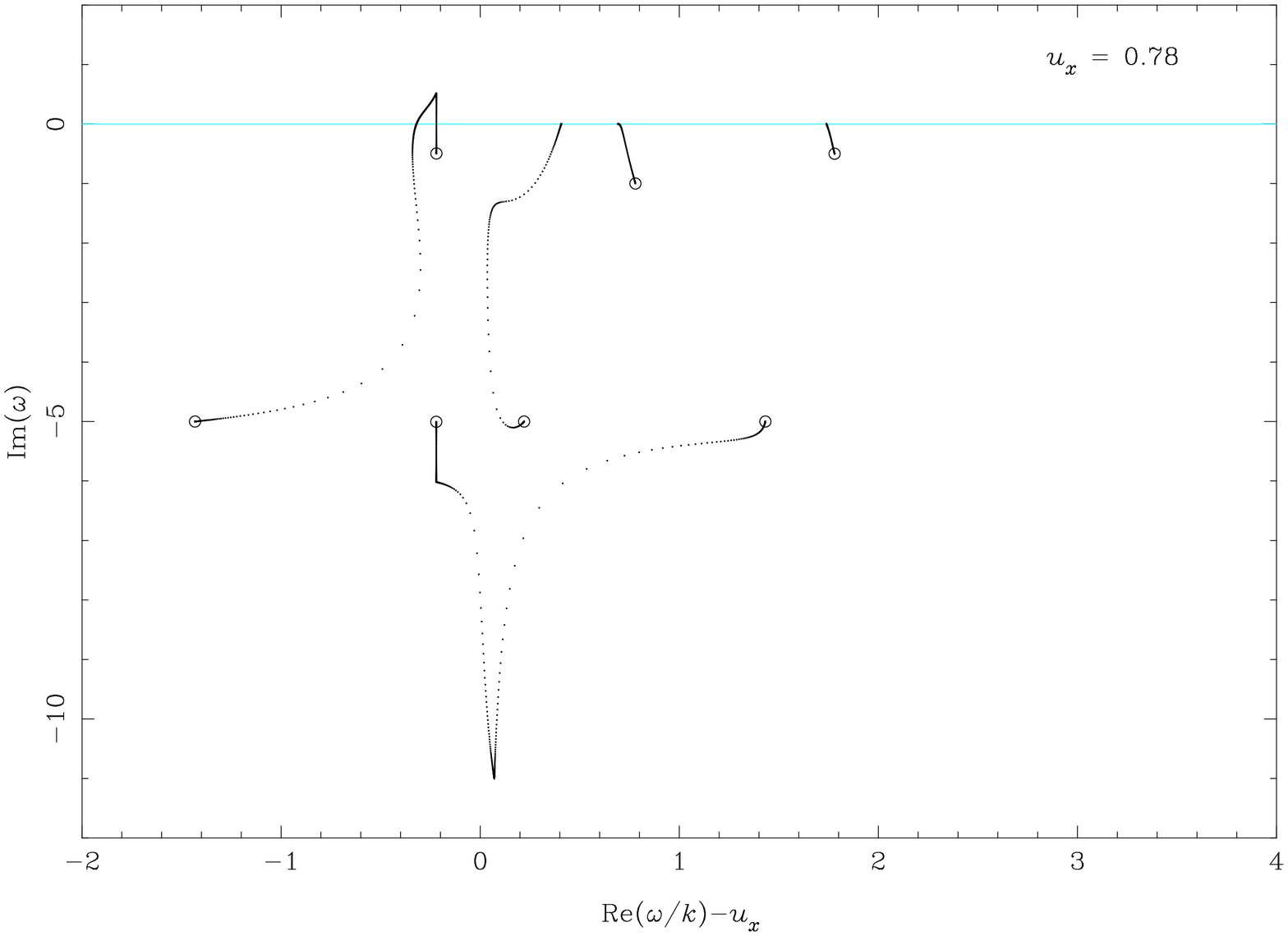}} &
\epsfysize=7cm\rotatebox{270}{\epsffile{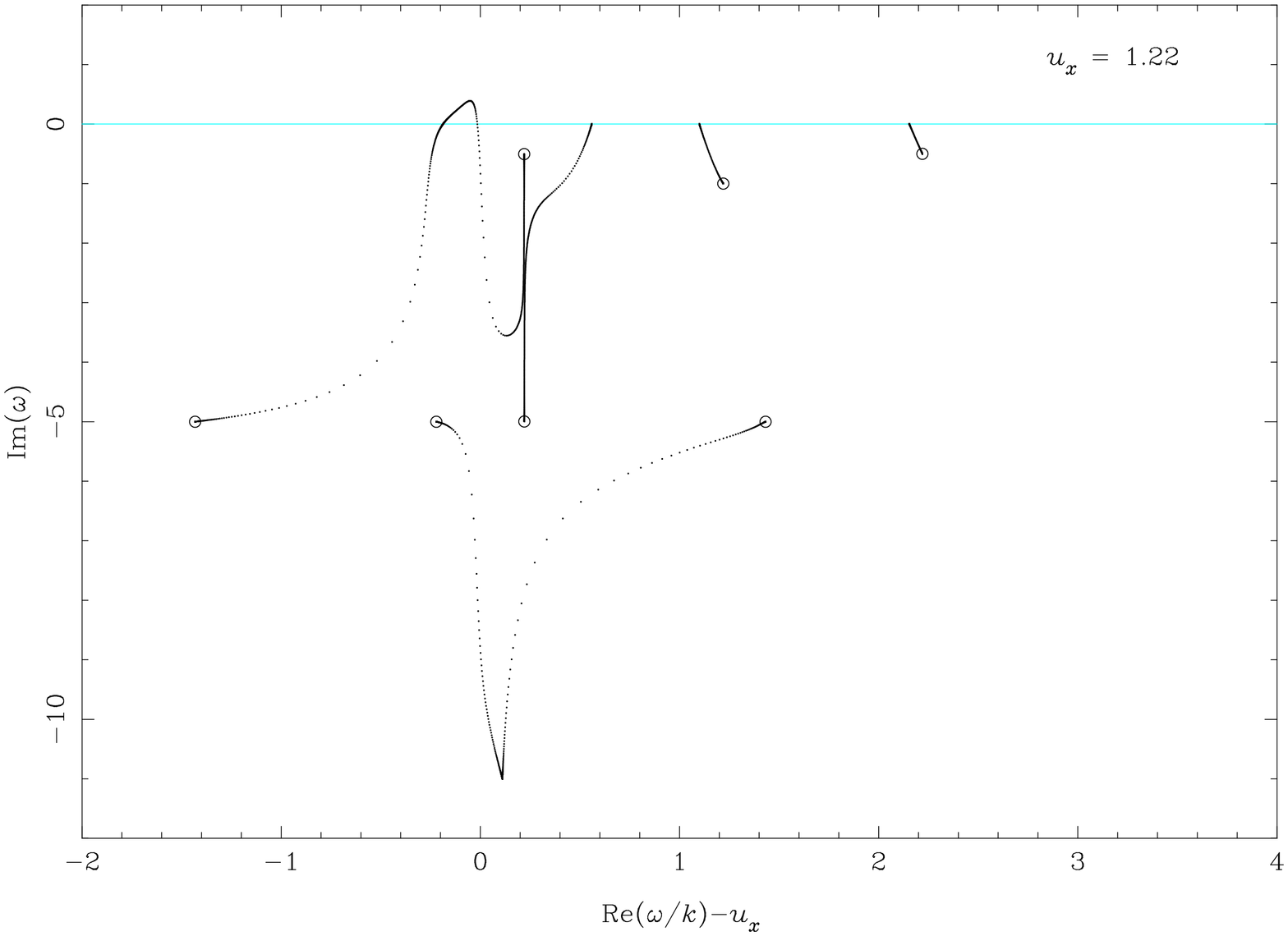}} \\
(f) & (g) \\
\epsfysize=7cm\rotatebox{270}{\epsffile{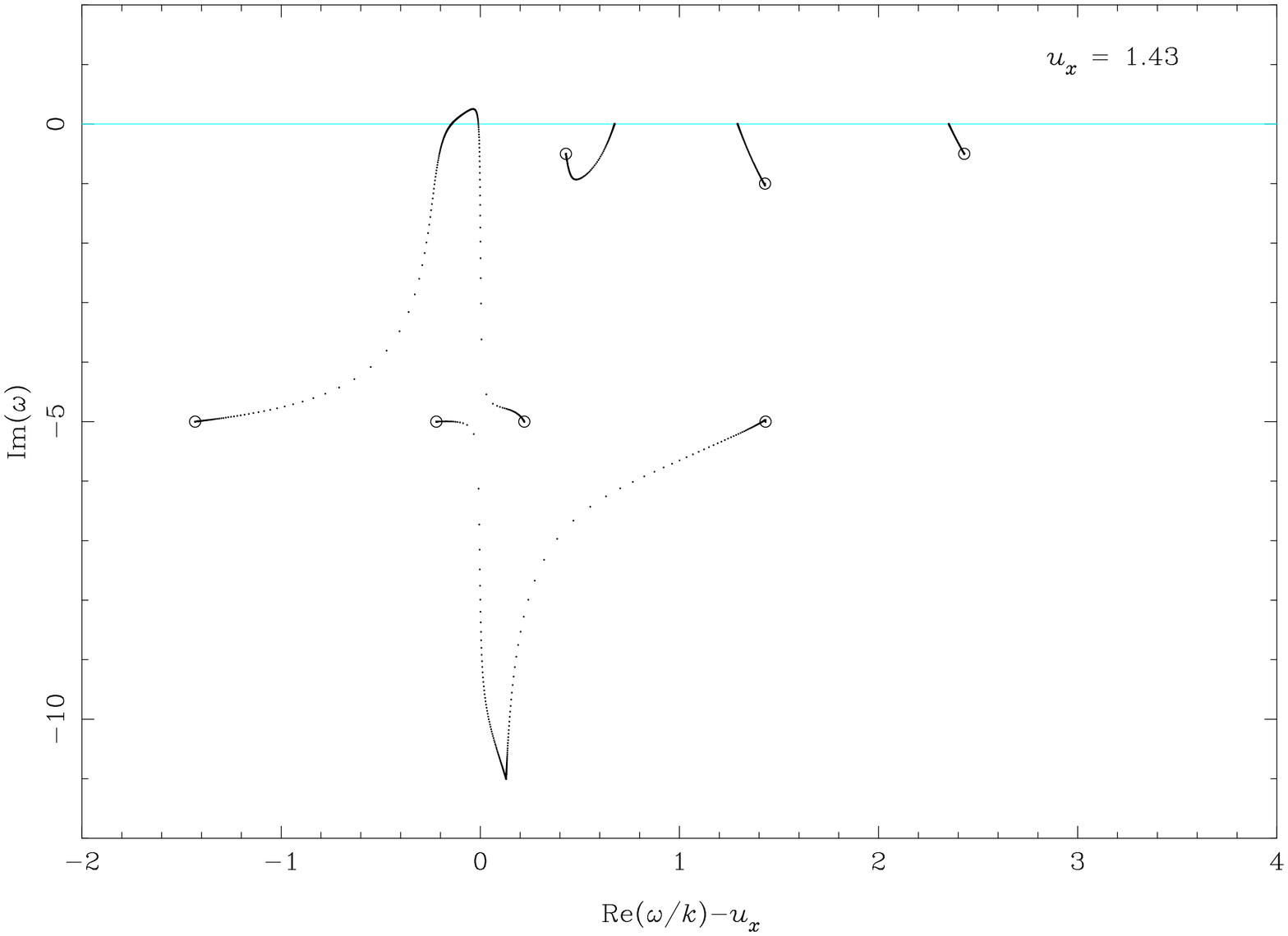}} &
\epsfysize=7cm\rotatebox{270}{\epsffile{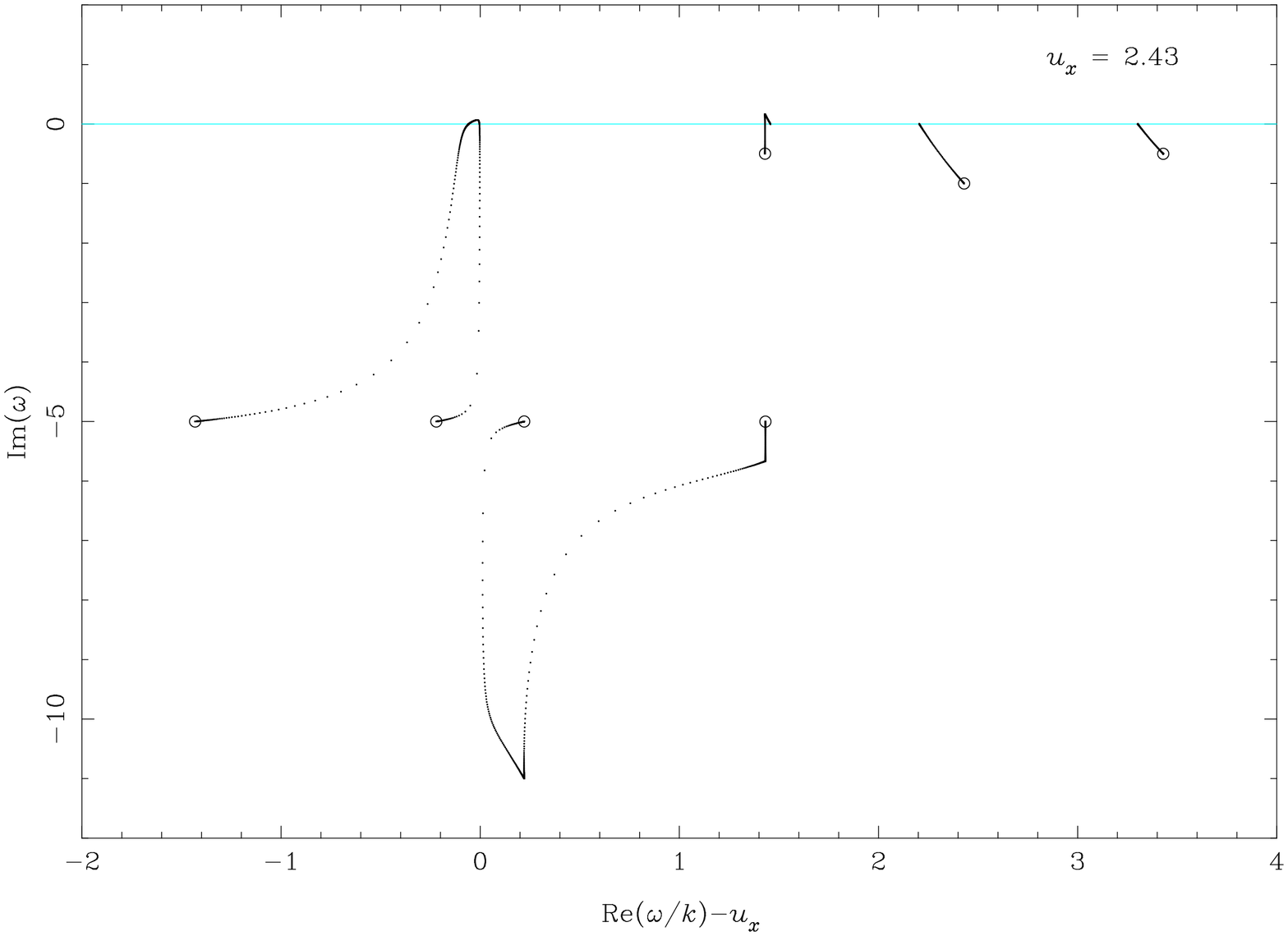}} \\
\end{tabular}
\end{centering}
\caption[]{Variation of eigenvalues of group B with $k$, for various
$u_x$.  As in Figure~\protect\ref{f:ex1}, $\bv{B} =
(1,0,1)\sqrt{\mu}$, $\lambda = 1$, $\rho = 1$, $\rho|n = 10$, $c|i^2 =
0.1$, $c|n^2 = 1$, $\bv{k} = (1,0,0)k$ and $\bv{u}=(1,0,-1)u_x$.  The
velocities $u_x$ are chosen to select various mode resonances.  The
circled points show the short-wavelength limiting values.}
\label{f:max1}
\end{figure*}

\begin{table}
\caption{Properties of the most unstable solutions in
Figure~\protect\ref{f:max1}.  The panels of this figure corresponding
to the various values of $u_x$ are noted.  Values are bracketed where
the most unstable modes are not continuous with the resonant modes.}\
\begin{center}
\begin{tabular}{cccc}
$u_x$ & Panel & ${\rm max}[{\rm Im}(\omega)]$ & 
	$k_x$ at ${\rm max}[{\rm Im}(\omega)]$ \\
0.22 & (b) & 0.020 & 33.1 \\
0.43 & (c) & (0.043) & (2.70) \\
0.78 & (d) & 0.516 & 173.8 \\
1.22 & (e) & (0.390) & (6.08) \\
1.43 & (f) & (0.253) & (4.06) \\
2.43 & (g) & 0.163 & 92.9
\end{tabular}
\end{center}
\label{t:max1}
\end{table}

In this section, we present numerical results for the roots of the
dispersion relations~(\ref{e:gpan}) and~(\ref{e:pdisp}), concentrating
in particular on circumstances where these roots correspond to
physical instabilities.  The roots were calculated as the eigenvalues
of the complex matrix corresponding to
equations~(\ref{e:lintn}--\ref{e:linbt}), using the routine
\verb!ZGEEVX! from LAPACK \cite{lapack}, and were verified by
comparison to direct solutions to the polynomial dispersion relations.
We show typical results in Figures~\ref{f:ex1} and~\ref{f:max1}.

Where the coupling is weak (i.e.\ at high wavenumbers) and when the
roots of the frozen system are well spaced, we find that the neutral
acoustic modes are damped as $\omega = \omega^0 - i\lambda\rho/2$, the
neutral shear modes as $\omega = \omega^0 - i\lambda\rho$ and the
ionized fast, slow and Alfv\'en modes all as $\omega = \omega^0 -
i\lambda\rho|n/2$, in agreement with the results in
Section~\ref{s:ndgen}.

Close to resonances, where two or more modes have similar phase
velocities, additional modes need to be taken into account.  The group
A modes never become unstable, as confirmed by the analysis of
Section~\ref{s:gpa}.  Instabilities are, however, found for group B
modes.  In Figure~\ref{f:ex1}, we show the values of
$\omega/k-\bv{u}_0.\hat{\bv{k}}$ for the group B modes, for one
particular set of values of $\lambda$, $\rho$, $c|i$, $\rho|n$, $c|n$
and $\bv{B}$, and direction $\hat{\bv{k}}$.  The plots in this figure
show the variation of the roots as the slip velocity is varied, while
the value of the wavenumber $k$ is increased through the set of plots.
Numerous graphs have been included to illustrate the full range of
topologies in phase space, and to allow the development of the
instabilities to be followed from linear through nonlinear order in
$\lambda$.

For large $k$ (or short wavelength), instabilities (where ${\rm
Im}(\omega)>0$) appear at resonances between the neutral sound waves
and the ionized fast- and slow-modes, and also (with a rather smaller
growth rate) at resonances between one neutral shear mode and the
ionized slow-modes.  As $k$ decreases, the morphology of the phase
space evolves, and additional modes begin to influence the
instabilities of the resonant system.  In particular, the roots
corresponding to the slow modes of the undamped system merge at $k
\sim 25$, leading to a cusped structure at ${\rm Re}(\omega)=0$ for
smaller $k$.  The apex of this cusp corresponds to solutions at $u_x
\to\pm\infty$.  This structure is particularly notable since, as $k$
decreases, the roots at increasingly large $u_x$ become unstable (cf.\
Section~\ref{longlimit}).  Nevertheless, comparing the panels of
Figure~\ref{f:ex1}, it is apparent that the unstable modes which are
seen can all be related to resonant instabilities for large $k$.

In a physical flow, waves with the full range wavenumbers $\bv{k}$
vary independently (at least in the linear limit), for fixed $\lambda$
and $\bv{u}$.  In Figure~\ref{f:max1}, we show the variation of the
roots for varying magnitude of the wave-vector, $k$, at fixed $u_x$
(i.e.\ each of the points shown on these graphs appear {\it
simultaneously}\/ in a single physical system, as each corresponds to
an independently-varying wave mode).  As $k$ decreases, each mode
moves from its value at $k\to\infty$ (shown circled), at which its
phase velocity is determined by the uncoupled equations.  As these
examples were all chosen with two wave modes close to resonance, the
resonant modes show a characteristic behaviour for large values of
$k$, diverging from their positions at $k\to\infty$ in a direction
essentially parallel to the imaginary axis.  In each case, as $k\to 0$
two of the solutions have ${\rm Im}(\omega)\to\lambda(\rho|i+\rho|n) =
11$, while the remaining five solutions have ${\rm Im}(\omega)\to 0$.
For intermediate $k$, at least one root is on occasion unstable for
each of the cases shown in Figure~\ref{f:max1}(b--g).  While each of
these cases corresponds to a different mode resonance, and it is clear
that these resonances have an important effect on the development of
the roots with decreasing $k$, only three of the resonances lead
directly to instabilities.

The properties of the most unstable modes in each of these cases are
given in Table~\ref{t:max1}: the cases in which the resonant modes
lead directly to instability have most unstable modes at the largest
wavenumbers.  That the other cases are unstable is a result of the
broad range of velocities away from resonance at which the three
unstable mode couplings lead to instability, rather than resulting
from the closer resonances which are present in these examples.
Analytic criteria corresponding to these results will be determined in
Section~\ref{s:twowave}.

We have shown results for each of the values of $k$ which are present
for particular physical parameters, for a particular direction of the
wavevector.  In the Section~\ref{s:geom}, we present a geometrical
argument which allows us to generalize these results for the full
range of wavevector directions present in a physical flow.

\section{Stability analysis}

\label{s:stability}
In this section, we will study the stability of the solutions to the
dispersion relations~(\ref{e:gpan}) and~(\ref{e:pdisp}) analytically.
First, we develop general stability criteria for dispersion relations
with the general form of equation~(\ref{e:pdisp}), including in
particular a discussion of the case in which two or more of the
undamped wave modes have similar frequencies.  We then apply these
results, and find that isolated modes are stable, and that the group A
modes remain so for all $\lambda$.  For the group B modes, however, we
find that several two-mode resonances lead to instability.  We also
study higher resonances, and the stability of modes in the
long-wavelength limit.

Our results confirm the importance of mode resonances for flow
stability, as observed in the numerical results of the previous
section.

\subsection{General considerations}

\label{s:general}

\subsubsection{Perturbation theory}

\label{s:pavres}
We first consider the limit in which the coupling parameter $\lambda$
is small, i.e., that the damping parameters $\Omega|i$ and $\Omega|n$
in the dispersion relations are small.  It is easy to see that
$P_0(\Omega)$ may be written
\begin{equation}
P_0(\Omega)=\prod_{\alpha}(\Omega-\Omega_{\alpha}^{(0)}),
\end{equation}
where $\Omega_{\alpha}^{(0)}=V_{\alpha}$ are eigenvalues of the
uncoupled system (i.e., when $\lambda=0$, and so
$\Omega|i=\Omega|n=0$).  We now use perturbation analysis, writing the
frequency of each mode of equation~(\ref{e:pdisp}) as a power-series
expansion in the parameter $\lambda$, which is assumed small:
\begin{equation}
\Omega_{\alpha}=\Omega_{\alpha}^{(0)}+\Omega_{\alpha}^{(1)}+\dots,
\end{equation}
where typically $\Omega_{\alpha}^{(n)}\propto\lambda^n$.

If there are no degenerate or near-degenerate roots in the uncoupled
system, then the first-order corrections may be written
\begin{equation}
\Omega_{\alpha}^{(1)}=-i\frac{\Omega|iP_1( \Omega_{\alpha}^{(0)})
+\Omega|n P_2(\Omega_{\alpha}^{(0)})}{\prod_{\beta,
\beta \neq \alpha}(\Omega_{\alpha}^{(0)}- \Omega_{\beta}^{(0)})}.
\label{e:pndgen}
\end{equation}
From this equation, it is apparent that for the flow to be stable, the
roots of $P_0(\Omega)$ and $\Omega|iP_1(\Omega)+\Omega|nP_2(\Omega)$
must interleave \cite[cf Appendix~\ref{a:hb};]{whitham}.  Since the
roots of $P_0(\Omega)$ interleave with those of $P_1(\Omega)$ and
$P_2(\Omega)$ individually, it is clear that, as $\Omega|i, \Omega|n >
0$, this is indeed the case.  (Note that this also demonstrates that
$\Omega|iP_1(\Omega)+\Omega|nP_2(\Omega)$ has six real roots, which
will be important in our discussion of long-wavelength modes below).

However, if there are two strictly degenerate modes of the uncoupled
system, with $\Omega=\Omega_{\alpha}^{(0)}$, then
equation~(\ref{e:pndgen}) can no longer be applied and we have to take
into account terms quadratic in $\lambda$ in our perturbation
analysis.  In this case, equations~(\ref{e:p0}--\ref{e:p22}) imply
\begin{equation}
P_1(\Omega_{\alpha}^{(0)}) = P_2(\Omega_{\alpha}^{(0)}) = 0.
\end{equation}
We then obtain the following quadratic equation for the first order
correction to the degenerate mode, $\Omega_{\alpha}^{(1)}$,
\begin{equation}
(\Omega_{\alpha}^{(1)})^2 \prod_{\beta, \Omega_{\beta}^{(0)} 
	\neq \Omega_{\alpha}^{(0)}} 
(\Omega_{\alpha}^{(0)}-\Omega_{\beta}^{(0)})+i\left[ \Omega|i 
P_1'(\Omega_{\alpha}^{(0)})+\Omega|n 
P_2'(\Omega_{\alpha}^{(0)}) \right] 
\Omega_{\alpha}^{(1)}-
\Omega|i^2 P_{11}(\Omega_{\alpha}^{(0)})
-\Omega|n^2 P_{22}(\Omega_{\alpha}^{(0)})-
\Omega|i \Omega|n P_{12}(\Omega_{\alpha}^{(0)})=0,\label{e:ndgen}
\end{equation}
where $P_1'(\Omega) = dP_1/d\Omega$.  This equation has two solutions,
which corresponds to the breaking of the degeneracy of the modes in
the presence of a small perturbation.  Note that even though the
expansion was to second order in quantities proportional to $\lambda$,
the correction is of first order.

If we have near-degenerate roots, then $V_{\alpha} \sim V_{\beta}$ for
some set of roots $\{\alpha,\beta,\dots\}$: we will refer to this
circumstance as a {\it resonance}.  As a result, we have to consider
the difference(s) $V_{\alpha}-V_{\beta}$, $\alpha \neq \beta$ as
further small parameter(s).  For a two-wave resonance, we have
\begin{eqnarray}
(\Omega_{\alpha}^{(1)})^2 \prod_{\gamma, \gamma \neq \{\alpha,\beta\}} 
(\Omega_{\alpha}^{(0)}-\Omega_{\gamma}^{(0)})+
\left[ \prod_{\gamma,\gamma\neq\alpha}
(\Omega_{\alpha}^{(0)}-\Omega_{\gamma}^{(0)})
+i\left(\Omega|i P_1'(\Omega_{\alpha}^{(0)})+
\Omega|n P_2'(\Omega_{\alpha}^{(0)})\right) \right] 
\Omega_{\alpha}^{(1)}+\nonumber\\
i\Omega|i P_1(\Omega_{\alpha}^{(0)})+
i\Omega|n P_2(\Omega_{\alpha}^{(0)})-
\Omega|i^2 P_{11}(\Omega_{\alpha}
^{(0)})-\Omega|n^2 P_{22}(\Omega_{\alpha}^{(0)})-
\Omega|i \Omega|n P_{12}(\Omega_{\alpha}^{(0)})&=&0,
\label{e:pavresv}
\end{eqnarray}
where the root of the quadratic with the smaller real part should be
chosen for continuity with the non-resonant case.

This equation can be written in the form
\begin{equation}
Q_2(\Omega_{\alpha}^{(1)})+iQ_1(\Omega_{\alpha}^{(1)})=0,
\label{e:Q}
\end{equation}
with
\begin{equation}
Q_2(\Omega_{\alpha}^{(1)})=(\Omega_{\alpha}^{(1)})^2 
\prod_{\gamma, \gamma \neq \{\alpha,\beta\}} 
(\Omega_{\alpha}^{(0)}-\Omega_{\gamma}^{(0)})+
\Omega_{\alpha}^{(1)} \prod_{\gamma,\gamma\neq\alpha}
(\Omega_{\alpha}^{(0)}-\Omega_{\gamma}^{(0)})-
\Omega|i^2 P_{11}(\Omega_{\alpha}
^{(0)})-\Omega|n^2 P_{22}(\Omega_{\alpha}^{(0)})-
\Omega|i \Omega|n P_{12}(\Omega_{\alpha}^{(0)})
\label{e:Q2}
\end{equation}
and
\begin{equation}
Q_1(\Omega_{\alpha}^{(1)})=\left(\Omega|i P_1'(\Omega_{\alpha}^{(0)})+
\Omega|n P_2'(\Omega_{\alpha}^{(0)})\right) 
\Omega_{\alpha}^{(1)}+\Omega|i P_1(\Omega_{\alpha}^{(0)})+
\Omega|n P_2(\Omega_{\alpha}^{(0)}).
\label{e:Q1}
\end{equation}

We can now apply the Hermite-Biehler theorem (see
Appendix~\ref{a:hb}), which tells us that the system described by the
equation
\begin{equation}
Q_n(\Omega_{\alpha}^{(1)})+iQ_{n-1}(\Omega_{\alpha}^{(1)})=0
\label{e:Qn}
\end{equation}
is stable (that is the roots of the equation have negative imaginary
parts) when the roots of the polynomials $Q_n$ and $Q_{n-1}$ are real
and interleave (and their leading terms are positive), and that if
this is not the case the system is unstable.  In the case of
equation~(\ref{e:Q}) we have polynomials of first and second degree,
and can easily obtain conditions for stability.  For higher
resonances, the expressions are clumsy, so it is better to find
instability conditions for particular cases rather than apply the
general expression.

\subsubsection{Geometry of resonances}
\label{s:geom}

\begin{figure*}
\begin{tabular}{ll}
(a) & (b) \\ \epsfysize=7cm \epsffile{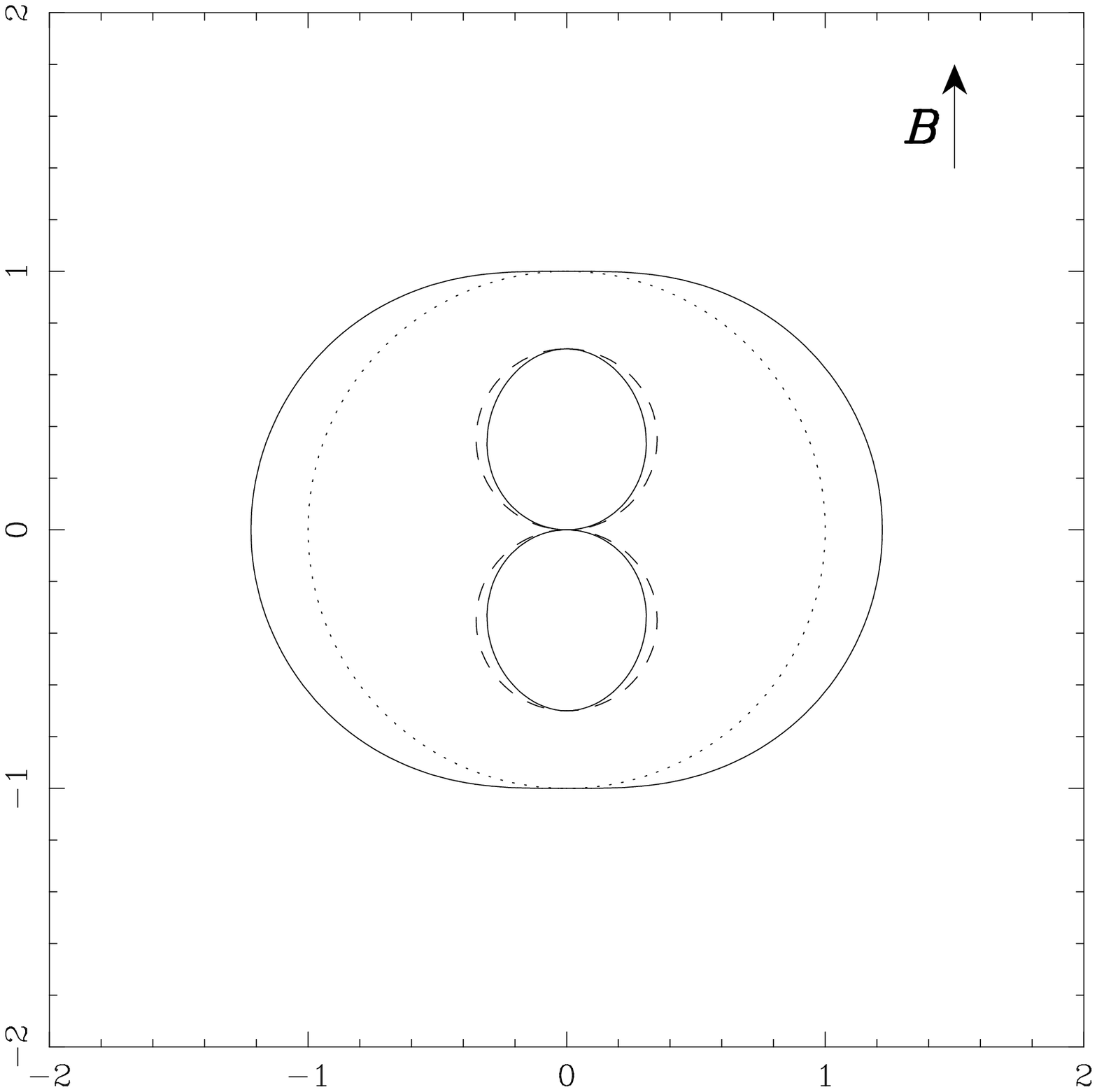} &
\epsfysize=7cm \epsffile{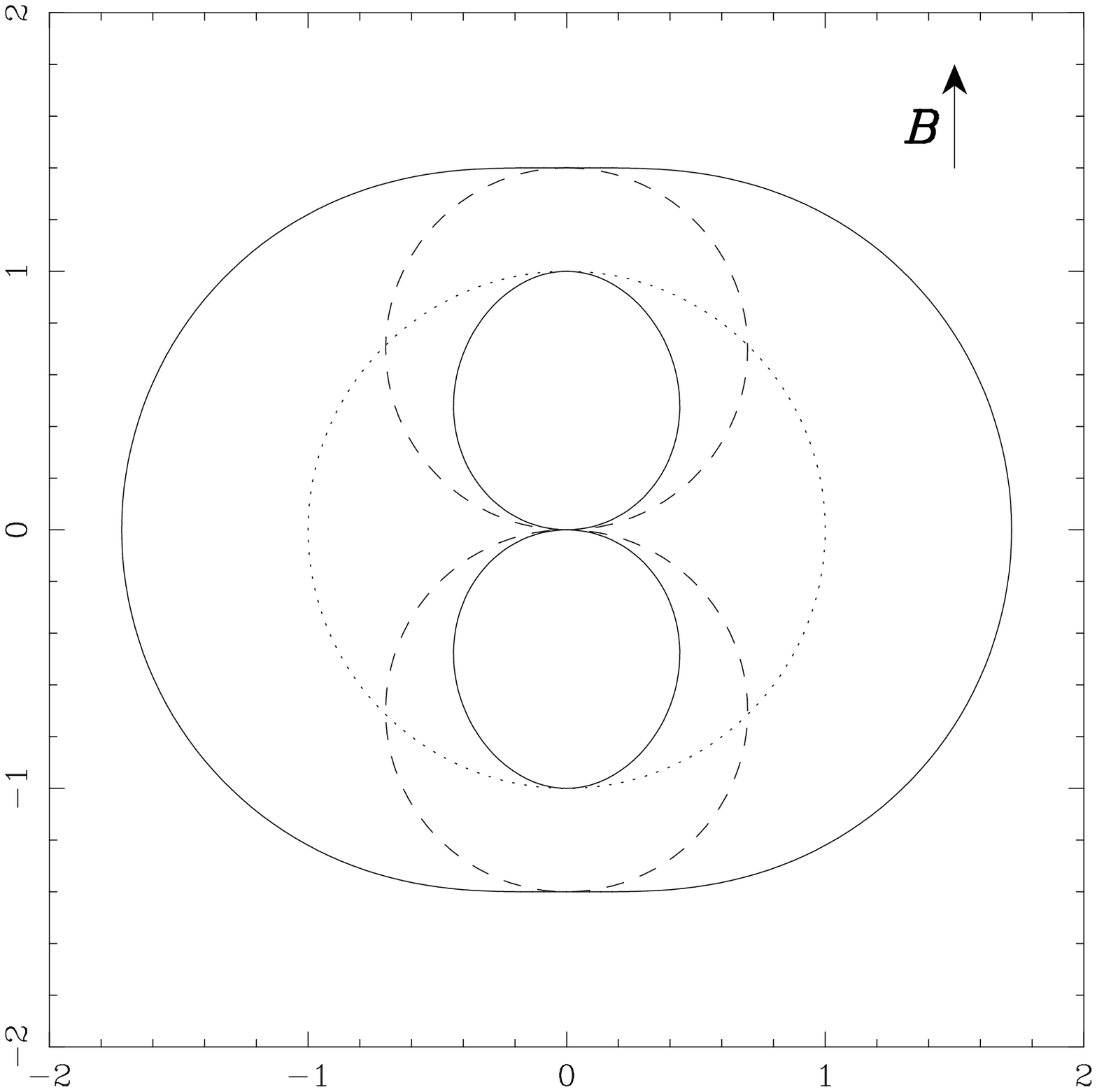} \\ (c) & (d) \\
\epsfysize=7cm \epsffile{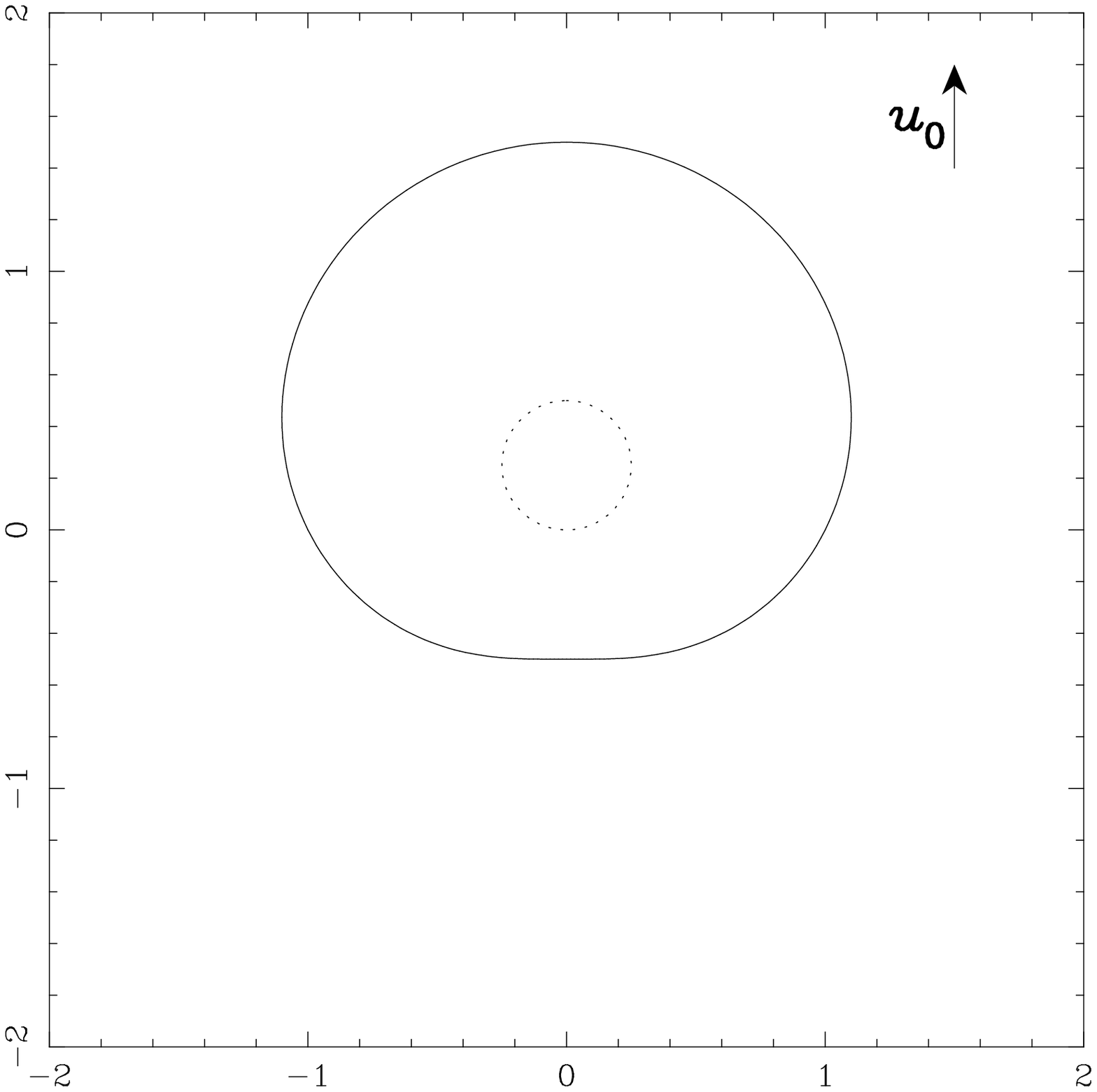} & \epsfysize=7cm
\epsffile{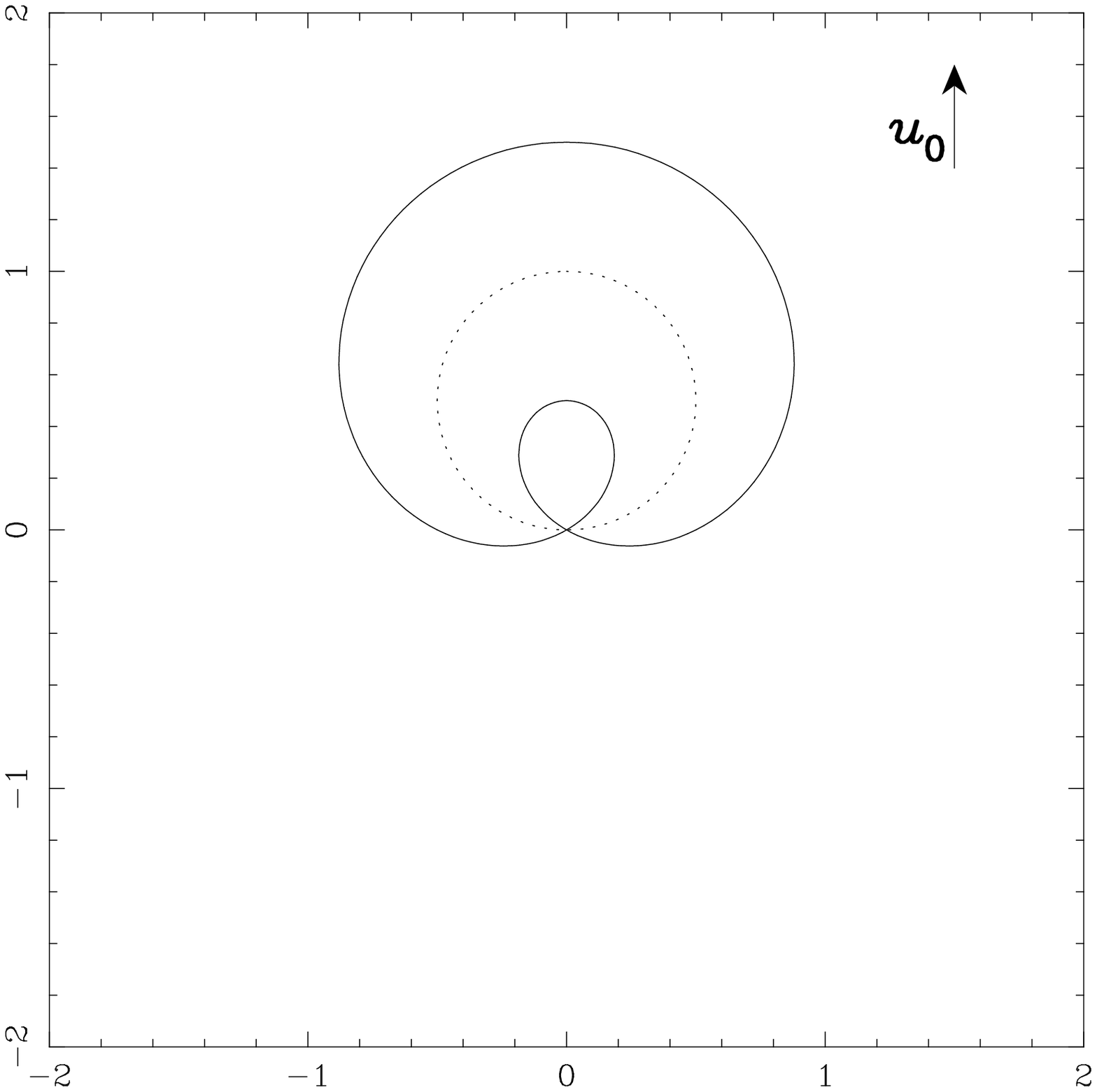} \\
\end{tabular}
\caption{Phase velocity profiles, $v = \omega/k$, as a function of the
angle between $\bv{k}$ and $\bv{B}$ or $\bv{u}_0$, plotted on polar
axes.  (a), (b) Solid - fast and slow, dashed - Alfv\'en, [the dotted
curve - sound - is included for comparison but has no dynamical role
here], $c|i > v|A$ for (a) and $c|i < v|A$ for (b); (c), (d) Solid -
sound, dotted - shear, with velocity offset, (c) is for subsonic flow
and (d) for supersonic flow.}
\label{f:crit}
\end{figure*}

In the preceding section, we presented a set of general criteria for
the stability of flows with internal damping parameters, and suggested
that mode resonances may play an important role.  We now find general
conditions for the presence of mode resonances in two-fluid MHD flows
in three dimensions, by comparing the geometry of phase diagrams for
the magnetosound modes of the ionized gas component and the sound and
shear modes of the neutral component.

In Figure~\ref{f:crit}(a) and (b), we show the phase speeds for the
fast, slow and Alfv\'en modes, as a polar plot of $\omega/k$ as a
function of the angle between $\bv{k}$ and the magnetic field
$\bv{B}$, while in Figure~\ref{f:crit}(c) and (d) we show the same for
sound waves and a shear mode in a frame in relative motion as a
function of the angle between $\bv{k}$ and $\bv{u}_0$.

It is clear that the passing of the phase curve for sound waves in
Figure~\ref{f:crit}(d) through the origin corresponds to the Cerenkov
condition for the emission of sound waves by an individual particle
moving at $\bv{u}_0$.  In a similar manner, resonance conditions
between modes in the different phases can identified by looking for
intersections between curves on suitably scaled and oriented versions
of these plots (remembering to keep the origin at the same place in
the ionized and the neutral plot).  For example, resonances will
always occur between the slow and shear modes, unless $\bv{u}_0$ is
parallel to $\bv{B}$ and $u_0>v|A$.  Likewise, resonances between the
slow and neutral sound modes will be present in general once the
relative motion is faster than the neutral sound speed, and for some
choices of parameters once the slip velocity is greater than the
difference between the neutral sound speed and the Alfv\'en velocity.

For finite damping, the resonance condition is weakened: the lines
shown in Figure~\ref{f:crit} may be thought of as blurred, although as
a side-effect additional modes must also be considered in the
stability analysis.

We have now determined general stability criteria, suggested the
importance of mode resonances in determining stability, and
demonstrated that they will be an almost universal feature of MHD
flows with drift velocities between components.  However, depending on
the form of the frictional force, only some of these resonances will
result in flow instabilities.  In the following subsections, we will
apply these results to the dispersion relations given in
Section~\ref{s:basic} to determine quantitative stability criteria.

\subsection{Non-resonant modes}

\label{s:ndgen}
Using relation~(\ref{e:pndgen}), first let us consider the stability
of the modes of equation~(\ref{e:pdisp}) when there are no resonances
in the uncoupled system, and $\Omega|i$ and $\Omega|n$ are small.  We
find that all the modes are stable.  In detail:

\begin{description}
\item[{\bf Neutral shear modes}]

For neutral shear modes, with $\Omega_{\alpha}^{(0)}=0$, 
\begin{equation}
\Omega_{\alpha}^{(1)}=-i\Omega|i.
\end{equation}

\item[{\bf Neutral sound waves}]

For neutral sound waves, with $\Omega_{\alpha}^{(0)}=\pm c|n$,
\begin{equation}
\Omega_{\alpha}^{(0)}=-i\frac{\Omega|i}{2}.
\end{equation}

\item[{\bf Magnetosound and Alfv\'en modes}]

For the fast and slow magnetosound waves, with
$\Omega_{\alpha}^{(0)}=u_x \pm v|{f,s}$, and also for the (group A)
Alfv\'en modes, with $\Omega_{\alpha}^{(0)}=u_x\pm v_{{\rm A}x}$,
\begin{equation}
\Omega_{\alpha}^{(1)}=-i\frac{\Omega|n}{2}.
\end{equation}

\end{description}

\subsection{Group A modes}

\label{s:gpa}
The stability of the group A solutions can easily be analysed for all
$\lambda$ (not necessarily small), by applying the Hermite-Biehler
theorem to the dispersion relation, equation~(\ref{e:gpan}).  This
equation couples two Alfv\'{e}n modes and one neutral shear mode, and
can be written
\begin{equation}
Q_3(\Omega)+iQ_2(\Omega)=0,
\end{equation}
where
\begin{equation}
Q_3(\Omega)=\left[ (\Omega-u_x)^2-v_{{\rm A}x}^2 \right] \Omega
\end{equation}
and
\begin{equation}
Q_2(\Omega)=\Omega|i \left[ (\Omega-u_x)^2-v_{{\rm A}x}^2 \right]
	+\Omega|n \Omega(\Omega-u_x).
\end{equation}
It is clear that both $Q_2$ and $Q_3$ have their full complement of
zeroes, and that the zeroes of $Q_2$ interleave with those of $Q_3$
(so long as there are no strict degeneracies and
$\Omega|i,\Omega|n>0$).  Hence, by the Hermite-Biehler theorem, the
group A modes are always stable.

\subsection{Two-wave resonances}

\label{s:twowave}
We now consider the stability of the flow close to resonances between
various pairs of group B modes.  In order to determine whether
instability occurs, we will follow the method of
section~\ref{s:pavres}.

\subsubsection{Ionic magnetosound-neutral sound resonance}

First, we consider the case of a resonance between a neutral sound
mode and an ionic magnetosound mode, so
\begin{equation}
\alpha c|n \sim u_x+\gamma v|{f,s},
\end{equation}
where $\alpha$ and $\gamma$ can independently take values $+1$ or
$-1$.  Applying the Hermite-Biehler theorem to the corresponding
equations (\ref{e:Q})--(\ref{e:Q1}), one finds that instability occurs
when the following inequalities hold:
\begin{equation}
v_{{\rm A}x} (v|{f,s}^2-v|{s,f}^2) \left[
-c|i v|{s,f}{\rm sgn}(v_{{\rm A}x})+
\gamma  (\alpha c|n v_{{\rm A}x}-\bv{v}|A.\bv{u}) \right] <0
\label{e:pav28}
\end{equation}
and
\begin{equation}
(\Omega|i+\Omega|n)^2>
-\frac{4v|{f,s} (v|{f,s}^2-v|{s,f}^2)/v_{{\rm A}x}}{
-c|iv|{s,f}{\rm sgn}(v_{{\rm A}x})+\gamma (\alpha 
c|n v_{{\rm A}x}-\bv{v}|A.\bv{u})}(u_x+\gamma v|{f,s}-
\alpha c|n)^2.\label{e:isdamp}
\end{equation}
In the limit of strict degeneracy, the r.h.s\@. of the second
condition goes to zero, so in this limit the condition is no
constraint.  Note also that equation~(\ref{e:pav28}) is the condition
that the r.h.s.  of equation~(\ref{e:isdamp}) is non-negative.
Equation~(\ref{e:isdamp}) is in effect a minimum condition on
$\lambda/k$ for instability to occur (implying instability for large
damping or long wavelength), for modes at a given distance from
resonance.  Note that these properties of conditions~(\ref{e:pav28})
and~(\ref{e:isdamp}) also apply to conditions~(\ref{e:pav38})
and~(\ref{e:isdamp2}) below.

We can now consider particular cases of these general relations.

\begin{enumerate}
\item $\alpha c|n \sim u_x+\gamma v|f$.

So long as $v_{{\rm A}x}\ne0$, equation (\ref{e:pav28}) may be written
\begin{equation}
\gamma v|f
\left[\alpha c|n -{\bv{v}|A.\bv{u}\over v_{{\rm A}x}}\right]<c|i^2. 
\end{equation}
If $\bv{v}|A \perp \bv{u}$ then
\begin{equation}
\gamma \alpha v|f  < {c|i^2\over c|n}.
\end{equation}

If $c|i=0$ as well, then
\begin{equation}
\gamma \alpha < 0,
\end{equation}
which is possible only if $\gamma \alpha=-1$, i.e.\ when the sound
and fast-mode waves are oppositely directed.

\item $\alpha c|n \sim u_x+\gamma v|s$.

Equation (\ref{e:pav28}) gives us
\begin{equation}
\gamma v|s
\left[\alpha c|n-{\bv{v}|A.\bv{u}\over v_{{\rm A}x}}\right] > c|i^2.
\end{equation}

If $\bv{v}|A \perp \bv{u}$ then
\begin{equation}
\gamma \alpha v|s > {c|i^2\over c|n}
\end{equation}
which is possible only if $\gamma \alpha=1$, i.e.\ where the phase
velocities of the sound and slow-mode waves are in the same direction.
Unlike the previous case, we cannot take $c|i=0$ in this relation,
because in that case the remaining slow-mode wave will also have a
similar frequency.  The resulting higher resonance will be considered
below.
\end{enumerate}

\subsubsection{Ionic magnetosound-neutral shear mode resonance}

We now consider the case when $u_x+\gamma v|{f,s} \sim 0$.  Applying
the Hermite-Biehler theorem to equations~(\ref{e:Q}--\ref{e:Q1}) gives
the instability conditions
\begin{equation}
\left[ v|{f,s}^3+\gamma v_{{\rm A}x}\bv{v}|A.\bv{u} \right]
(v|{f,s}^2-v|{s,f}^2) <0 \label{e:pav38}
\end{equation}
and
\begin{equation}
(2\Omega|i+\Omega|n)^2>-\frac{4v|{f,s}(v|{f,s}^2-v|{s,f}^2)^2/v|A^2
}{v|{f,s}^3+\gamma v_{{\rm A}x}\bv{v}|A.\bv{u}}(u_x+\gamma v|{f,s})^2.\label{e:isdamp2}
\end{equation}

As before, we consider two particular cases

\begin{enumerate}
\item $u_x+\gamma v|f \sim 0$.

In this case equation (\ref{e:pav38}) gives us
\begin{equation}
v|f^3<-\gamma v_{{\rm A}x} \bv{v}|A.\bv{u}
\end{equation}
If $\bv{v}|A \perp \bv{u}$ then the interacting modes are always stable.

\item $u_x+\gamma v|s \sim 0$

Now we have from (\ref{e:pav38})
\begin{equation}
v|s^3>-\gamma v_{{\rm A}x} \bv{v}|A.\bv{u}\label{e:snlim}
\end{equation}
so that if $\bv{v}|A \perp \bv{u}$ then this mode interaction will
always lead to instability for sufficiently large $\lambda$, so long
as no other modes start to interact.
\end{enumerate}

\subsection{Degenerate slow magnetosound waves}

\label{s:ceren}
If we have $c|i \sim 0$ (which is often the case in astrophysical
applications), then the two slow magnetosound waves are resonant,
i.e.\ $u_x-v|s \sim u_x+v|s$ because $v|s \sim 0$.  This resonance is
of a different nature to those discussed in Section~\ref{s:twowave}, as
it is between two modes which propagate in the {\it same}\/ phase.

When $c|i$ is strictly zero, two slow magnetosound waves become
degenerate with common eigenvalue $\Omega_{\alpha}^{(0)}=u_x$.  In
order to investigate stability in this case, we have to solve equation
(\ref{e:ndgen}), and find
\begin{equation}
\Omega_{\alpha}^{(1)}=-i\frac{\Omega|n}{2} \pm \frac{i}{2} \sqrt{\Omega|n^2+
\frac{4\Omega|i\Omega|nc|n^2 v_{{\rm A}x}
	(v_{{\rm A}x}u_x-\bv{v}|A.\bv{u})}{u_xv|A^2(u_x^2-c|n^2)}}.
		\label{e:twores}
\end{equation}

Instability can occur for the mode with positive choice of sign if
\begin{equation}
v_{{\rm A}x} u_x (u_x^2-c|n^2)[v_{{\rm A}x}u_x-\bv{v}|A.\bv{u}]>0,
\label{e:concorde}
\end{equation}
For small but finite $c|i$, for instability to occur then in addition
to condition~(\ref{e:concorde}), we also require that
\begin{equation}
\Omega|i \Omega|n>-\frac{u_x (c|n^2-u_x^2) v|s^2v|A^2}{c|n^2
v_{{\rm A}x}[v_{{\rm A}x}u_x-\bv{v}|A.\bv{u}]}.
\end{equation}

For the important case $\bv{v}|A \perp \bv{u}$,
condition~(\ref{e:concorde}) leads to the simple relation
\begin{equation}
u_x^2>c|n^2,
\end{equation}
so long as $v_{{\rm A}x}\ne 0$.  Note that the perturbation analysis
will not apply when $u_x^2$ is close to $c|n^2$, since
$\Omega_{\alpha}^{(1)}$ given by equation~(\ref{e:twores}) will be no
longer be small.  While no resonance with modes in the neutral phase
has been assumed, this condition clearly relates to a resonance effect
with the neutral sound waves in some fashion (indeed, when followed in
numerical solutions, the mode is continuous with instabilities which
originate at the slow magnetosound/neutral sound resonance for finite
$c|i$).  As the ionized sound speed becomes small, the system of
coupled slow magnetosound modes becomes sensitive to the presence of
other modes, even those rather far from apparent resonance conditions
\cite[cf.\ the sensitivity of the slow-mode waves to driving by other
modes of nonlinear amplitude in the ionized gas,]{fh02}.

\subsection{Higher resonances}

We now consider the stability of some higher order mode couplings.

\subsubsection{Ionic slow magnetosound-neutral shear mode resonance}
\label{s:hr}

Let us consider the case when $u_x-v|s \sim 0 \sim u_x+v|s$, which is
possible if $c|i$ and $u_x$ are small.

First let us investigate the case $c|i=u_x=0$, in which when we
have a three times degenerate root $\Omega_{\alpha}^{(0)}=0$ for the
uncoupled system.  If we apply perturbation analysis to these roots,
we obtain the following equation for the correction
$\Omega_{\alpha}^{(1)}$:
\begin{equation}
(\Omega_{\alpha}^{(1)})^3+i(\Omega|i + \Omega|n) (\Omega_{\alpha}^{(1)})^2-
\Omega|i \Omega|n {v_{{\rm A}x} [\Omega_{\alpha}^{(1)}v_{{\rm A}x}-\bv{v}|A.
\bv{u}]\over v|A^2}=0.
\end{equation}
The polynomial $Q_2(\Omega_{\alpha}^{(1)})$ has two identical roots
$\Omega_{\alpha}^{(1)}=0$, which means that in order to satisfy the
interleaving criterion, the polynomial $Q_3(\Omega_{\alpha}^{(1)})$
must also have one root $\Omega_{\alpha}^{(1)}=0$.  This is possible
only if $\bv{v}|A.\bv{u}=0$ -- indeed for
\begin{equation}
\vert \bv{v}|A.\bv{u}\vert > 2\sqrt{\Omega|i\Omega|n\over 27} {v_{{\rm A}x}^2\over v|{A}}
\end{equation}
$Q_3$ has only a single real root.  From this we conclude that the
system is marginally stable when $\bv{v}|A.\bv{u}=0$ and unstable when
$\bv{v}|A.\bv{u} \neq 0$.  In the latter case, the equation has one
term of second order, so the roots are
\begin{equation}
\Omega_{\alpha}^{(1)} = \left(-1,{1\pm\sqrt{3}i\over2}\right)
\left(\Omega|i\Omega|n v_{{\rm A}x}\bv{v}|A.\bv{u}\over v|A^2\right)^{1/3},
\label{e:23sol}
\end{equation}
which scale as $\lambda^{2/3}$ for $\lambda\to 0$, with one root being
unstable.

Now we consider the general resonant case where $u_x$ and $c|i$
are small but finite.  For the $\Omega_{\alpha}^{(0)}=0$ mode we have
\begin{equation}
\Omega_{\alpha}^{(1)} \left[ (\Omega_{\alpha}^{(1)}-u_x)^2-v|s^2 \right]+
i\Omega|i \left[ (\Omega_{\alpha}^{(1)}-u_x)^2-v|s^2 \right] + i\Omega|n
\Omega_{\alpha}^{(1)} (\Omega_{\alpha}^{(1)}-u_x)-\Omega|i \Omega|n
{v_{{\rm A}x} [\Omega_{\alpha}^{(1)} v_{{\rm A}x}-\bv{v}|A.\bv{u}]\over 
	v|A^2}=0.
\label{e:three}
\end{equation}
The corrections derived for the roots which have
$\Omega_{\alpha}^{(0)}=u_x+\gamma v|s$ give equivalent results,
relative to their alternative datum, as must be the case since the
three roots of equation~(\ref{e:three}) correspond to the three
resonant modes.

The $Q_2$ polynomial for equation~(\ref{e:three}) has two real roots
\begin{equation}
\Omega_{\alpha}^{(1)}=\frac{(2\Omega|i+\Omega|n)u_x \pm 
\sqrt{\Omega|n^2 u_x^2+4\Omega|i (\Omega|i+\Omega|n)v|s^2}}
{2(\Omega|i+\Omega|n)}.
\label{roots}
\end{equation}

For the $Q_3$ polynomial, we first treat the case $\bv{v}|A.\bv{u}=0$.
Here, there are three real roots, one of which is zero and other two
are
\begin{equation}
\Omega_{\alpha}^{(1)}=u_x \pm \sqrt{v|s^2+\Omega|i \Omega|n 
	{v_{{\rm A}x}^2\over v|A^2}}.
\end{equation}
Interleaving of these roots with those given in equation~(\ref{roots})
is violated (that is, the system is unstable) when
\begin{equation}
v|s < \vert u_x \vert
	< \frac{v|s^2+\Omega|i (\Omega|i +\Omega|n) 
(v_{{\rm A}x}^2/v|A^2)}{\sqrt{v|s^2+ 
	\Omega|i \Omega|n (v_{{\rm A}x}^2/v|A^2)}}, 
\label{cond}
\end{equation}
so we see that the system, which is marginally stable under fully
degenerate conditions, can be unstable under resonant conditions.  The
range of unstable conditions widens in velocity as $\Omega|i$ and
$\Omega|n$ increase.

If $\bv{v}|A.\bv{u} \neq 0$, we have an equation of third degree for
$Q_3$,
\begin{equation}
\Omega_{\alpha}^{(1)} (\Omega_{\alpha}^{(1)}-u_x)^2-
\left(v|s^2 +\Omega|i \Omega|n {v_{{\rm A}x}^2\over v|A^2}\right) 
\Omega_{\alpha}^{(1)} +\Omega|i \Omega|n
{v_{{\rm A}x} \bv{v}|A.\bv{u}\over v|A^2}=0.
\label{eq1}
\end{equation}
Again the last term in this equation is of second order, so in the
short-wavelength limit the roots are still given by
equation~(\ref{e:23sol}).

In the more general case where $\Omega|i\Omega|n \ga
v|A^2(\bv{v}|A.\bv{u})^2/v_{{\rm A}x}^4$, equation~(\ref{eq1}) has
complex roots (so the system is unstable) if
\begin{equation}
\left(2u_x^3-18\Gamma_1 u_x+27\Gamma_2\right)^2
> 4\left(u_x^2+3\Gamma_1\right)^3
\label{uau}
\end{equation}
\cite[from the usual relations for cubic equations, \eg{}]{nr}, where
we have introduced new variables
\begin{eqnarray}
\Gamma_1 &=& v|s^2 +\Omega|i \Omega|n {v_{{\rm A}x}^2\over v|A^2}\\
\Gamma_2 &=& \Omega|i \Omega|n {v_{{\rm A}x} \bv{v}|A.\bv{u}\over v|A^2}.
\end{eqnarray}
This implies that the solutions are unstable as
$\bv{v}|A.\bv{u}\to\pm\infty$, other terms being equal: in particular,
if $u_x$ is zero there are complex roots when
\begin{equation}
\left\vert\Gamma_2\right\vert > 2\left(\Gamma_1\over3\right)^{3/2}.
\end{equation}
If $\vert\Gamma_2\vert \ll \Gamma_1^{3/2}$, condition~(\ref{uau}) is
satisfied for $u_x$ in the range
\begin{equation}
-{\rm sgn}(\Gamma_2)\Gamma_1^{1/2} - \left(2\vert\Gamma_2\vert\over
\Gamma_1^{1/2}\right)^{1/2} <u_x< -{\rm sgn}(\Gamma_2)\Gamma_1^{1/2} +
\left(2\vert\Gamma_2\vert\over \Gamma_1^{1/2}\right)^{1/2}.
\label{e:smallg2}
\end{equation}

If inequality~(\ref{uau}) is violated, then equation~(\ref{eq1}) has
three real roots and we have to solve the problem of interleaving.
Criteria for interleaving may be obtained by inserting the roots of
the $Q_2$ polynomial, equation~(\ref{roots}), into the l.h.s. of
equation~(\ref{eq1}).  These are rather complex criteria to apply,
except numerically.  In general, however, for small $\Omega|i$,
$\Omega|n$ the flow is unstable for conditions along whichever of the
lines $u_x = \pm v|s$ satisfies equation~(\ref{e:snlim}); as
$\Omega|i$, $\Omega|n$ increase this region of instability broadens
and also expands along the other of $u_x = \pm v|s$, in agreement with
the limiting cases given by equations~(\ref{cond})
and~(\ref{e:smallg2}).

\subsubsection{Neutral sound-neutral shear mode resonance}

Let us consider the case $\alpha c|n \sim 0$. One can show that
equation for the corrections has the form
\begin{equation}
\Omega_{\alpha}^{(1)}\left[ (\Omega_{\alpha}^{(1)})^2-c|n^2 \right]+
i\Omega|i \left[ 2(\Omega_{\alpha}^{(1)})^2-c|n^2 \right] -\Omega|i^2
\Omega_{\alpha}^{(1)}=0.
\end{equation}
It is obvious that roots of $Q_3$ and $Q_2$ always interleave, and so
the system is always stable under these resonant conditions (as is
clear physically).

\subsubsection{Five-wave resonance}

There is also the possibility of a five-wave resonance if $0 \sim
\alpha c|n \sim u_x+\gamma v|s$.  But this case gives us a fifth order
polynomial, which would be extremely complicated to analyze.  So here
we restrict ourselves to investigating only the case of full
degeneracy, e.g. $c|n=0$, $c|i=0$ and $u_x=0$.  Then we have three
roots $\Omega=0$, and for the remaining two we have the equation for
the corrections
\begin{equation}
(\Omega_{\alpha}^{(1)})^2+i\Omega_{\alpha}^{(1)} (2\Omega|i+\Omega|n)-
\Omega|i(\Omega|i+\Omega|n)=0.
\end{equation}
It is clear from this that the system is stable. However, as we have
seen above, the marginal stability of the three roots $\Omega=0$ under
degenerate conditions does not guarantee stability under resonant
conditions.

\subsection{Long-wavelength stability}
\label{longlimit}

We now consider the stability of modes in long wave limit, i.e.\ for
small wave-numbers.  Note first that if we take the limit $k
\rightarrow 0$ in equation (\ref{e:gpb}) then five modes have
behaviour $\omega \rightarrow O(k)$ and two modes have behaviour
$\omega \rightarrow -i \lambda (\rho|i+\rho|n)+O(k)$ (cf.\ also the
behaviour of the roots for long wavelengths shown in
Figures~\ref{f:ex1}(l) and~\ref{f:max1}).

Now we have to find the corrections to these limits of order $k$:
\begin{equation}
\omega= V_\alpha k, \;  \alpha=1 \dots 5
\label{firom}
\end{equation}
and
\begin{equation}
\omega=-i\lambda(\rho|i+\rho|n)+V_\alpha k, \;   \alpha=6,7.
\label{secom}
\end{equation}

According to the definition of the limit, small $k$ means that the
second term in equation~(\ref{secom}) is much smaller then the first
one.  As the first term has a negative imaginary part, the smaller
second term cannot change the overall sign.  Hence the roots given by
equation~(\ref{secom}) are stable in long wavelength limit.

It is more difficult to determine the stability of the roots given by
equation (\ref{firom}).  To first order in $k$, the stability of these
five modes in the limit of long wavelengths is determined by the roots
of
\begin{equation}
\Omega|i^2P_{11}(\Omega)+\Omega|i\Omega|nP_{12}
	(\Omega)+\Omega|n^2P_{22}(\Omega)=0,
\label{e:long}
\end{equation}
together with the condition that the leading term in this equation has
a positive coefficient (which is the case). This equation is
independent of $k$, and its roots are the velocities $V_\alpha$ in
equation~(\ref{firom}).  As equation~(\ref{e:long}) is a polynomial
equation with real coefficients, complex roots can arise as conjugate
pairs and any such complex roots in equation~(\ref{e:long}) will
result in instability in the limit of small $k$.

Real roots of equation~(\ref{e:long}) are marginally stable to first
order.  To determine their asymptotic stability, we would need to
study the interleaving of these roots with those of $\Omega|i
P_1(\Omega)+\Omega|nP_2(\Omega)$: if the roots of these two
polynomials do not interleave, the asymptotic solution will be
unstable at order $k^2$.  Note that we know that $\Omega|i
P_1(\Omega)+\Omega|nP_2(\Omega)$ has its full complement of real
roots, from the discussion after equation~(\ref{e:pndgen}) above. 

We will first study the stability of the long-wavelength solutions in
various limiting cases.  For ${\bf u}=0$ we have five real (and hence
stable) roots, which correspond to the fast- and slow-mode waves of
the fully coupled system together with an additional mode with
$\Omega=0$, as would be expected on physical grounds.  

If $u_x$ is large compared to any other characteristic velocity, the
solutions are $\Omega = 0$ [or more accurately
$\Omega|nc|n^2/(\Omega|i u_x)$], and $\Omega|iu_x/(\Omega|i+\Omega|n)$
and $u_x$ (the latter two are double roots).  This can be easily seen
if we write equation (\ref{e:long}) as follows
\begin{eqnarray}
\Omega (\Omega-u_x)^2
\left[(\Omega|i+\Omega|n)\Omega-\Omega|iu_x \right]^2-(\Omega-u_x)
\left[(\Omega|i+\Omega|n)\Omega-\Omega|iu_x \right]
\left[\Omega|i(v|A^2+c|i^2)\Omega+\Omega|n
c|n^2(\Omega-u_x)\right]+& & \nonumber \\
\Omega|i^2c|i^2v_{{\rm A}x}^2\Omega+\Omega|i \Omega|n c|n^2
v_{{\rm A}x} \left[v_{{\rm A}x}\Omega-{\bf v}|A.{\bf u} \right]&=&0.
\label{e:long1}
\end{eqnarray}
The presence of double roots means these solutions are only marginally
stable in the limit, so we need to carry the analysis to higher order
to determine stability in the asymptotic regime.  To lift these
degeneracies, we should expand to lower order in $\Omega$ and $u_x$,
or to higher order in $k$ (or include nonlinear terms neglected in our
initial linearization of the problem).

If ${\bf k} \perp {\bf v}|A$, i.e. $v_{{\rm A}x}=0$ with all other
terms finite, equation (\ref{e:long1}) has roots $\Omega=u_x$ and
$\Omega=\Omega|i u_x/(\Omega|i+\Omega|n)$.  The remaining equation of
three roots satisfy
\begin{equation}
\alpha|i\Omega\left[(\Omega-u_x)^2-(v|A^2+c|i^2)\right]
+\alpha|n(\Omega-u_x)(\Omega^2-c|n^2) = 0,\label{e:longnovax}
\end{equation}
where $\alpha|i=\Omega|i/(\Omega|i+\Omega|n)$,
$\alpha|n=\Omega|n/(\Omega|i+\Omega|n)$.  Considering the ordering of
the roots of the factor proportional to $\alpha|i$ and that
proportional to $\alpha|n\equiv 1-\alpha|i$, it is clear that there
must be at least two real roots of equation~(\ref{e:longnovax}) for
all $0<\alpha|i<1$, and hence there will be three real roots.
Therefore for $v_{{\rm A}x}=0$, all the solutions are stable to order
$O(k)$.

If we consider corrections to the $O(u_x)$ asymptotic solutions, then
for the two roots $\Omega = u_x+\delta$, we find that the highest
order corrections have $\delta = O(1/u_x)$, and are given by
\begin{equation}
\delta = {\Omega|i(v|A^2+c|i^2)\over 2 \Omega|n u_x}
\left[1\pm\sqrt{1+4{\Omega|n c|n^2 v_{{\rm A}x}{\bf v}|A.{\bf u}/u_x
	- v_{{\rm A}x}^2(\Omega|nc|n^2+\Omega|ic|i^2)
	\over
	\Omega|i(v|A^2+c|i^2)^2}}\right],
\end{equation}
which can clearly be complex, e.g. if ${\bf v}|A.{\bf u}=0$ and $c|i =
0$ the criterion for instability becomes $4v_{{\rm A}x}^2c|n^2/v|A^4
>\Omega|i/\Omega|n$.  Similar expressions for the case $\Omega \simeq
\Omega|iu_x/(\Omega|i+\Omega|n)$ are more complex, but the condition
for instability is 
\begin{equation}
4\Omega|i^2\left[v_{{\rm A}x}^2\Omega|i(\Omega|ic|i^2+\Omega|nc|n^2)
-\Omega|n(\Omega|i+\Omega|n)c|n^2v_{{\rm A}x}{\bf v}|A.{\bf u}/u_x\right]
> [\Omega|i^2(v|A^2+c|i^2)+\Omega|n^2c|n^2]^2.
\end{equation}

If, instead, we include higher order terms in $k$, where in the limit
of large $u_x$
\begin{eqnarray}
P_0 &=& \Omega^3(\Omega-u_x)^4,\\
\Omega|iP_1+\Omega|nP_2 &=& 2\Omega^2(\Omega-u_x)^3
	[(\Omega|i+\Omega|n)\Omega-u_x\Omega|i],\\
\Omega|i^2P_{11}+\Omega|i\Omega|nP_{12}+\Omega|n^2P_{22}&=& 
	\Omega(\Omega-u_x)^2
	[(\Omega|i+\Omega|n)\Omega-u_x\Omega|i]^2,
\end{eqnarray}
then the common roots $\Omega = 0$ and $u_x$ (twice) are marginally
stable.  Interleaving is satisfied for the remaining four roots, so
they are each stable for all $k$ at large $u_x$ (although only the
limiting form as $k\to0$ is required for the present discussion).

First-order instabilities can also arise in the intermediate velocity
regime.  Looking again at equation~(\ref{e:long1}), we have shown that
quintic consisting of the first two terms has five real roots.
However, the values $v_{{\rm A}x}$ and $\bv{v}|A.\bv{u}$ can be varied
independently of any of the other parameters of the overall equation,
so the additional linear function consisting of the third and fourth
terms is entirely arbitrary.  Even when $\bv{v}|A.\bv{u}=0$, the
number of real roots of equation~(\ref{e:long1}) can be as few as one.

As the long wave limit is quite important for the astrophysics of
molecular clouds and stellar winds, in the next section we will study
a typical astrophysical example, SiO maser spots in late-type stellar
winds.

\section{SiO maser spots in late-type stellar winds}
\label{s:wind}

\begin{figure*}
\begin{centering}
\begin{tabular}{ll}
(a) & (b) \\
\epsfysize=8cm\rotatebox{270}{\epsffile{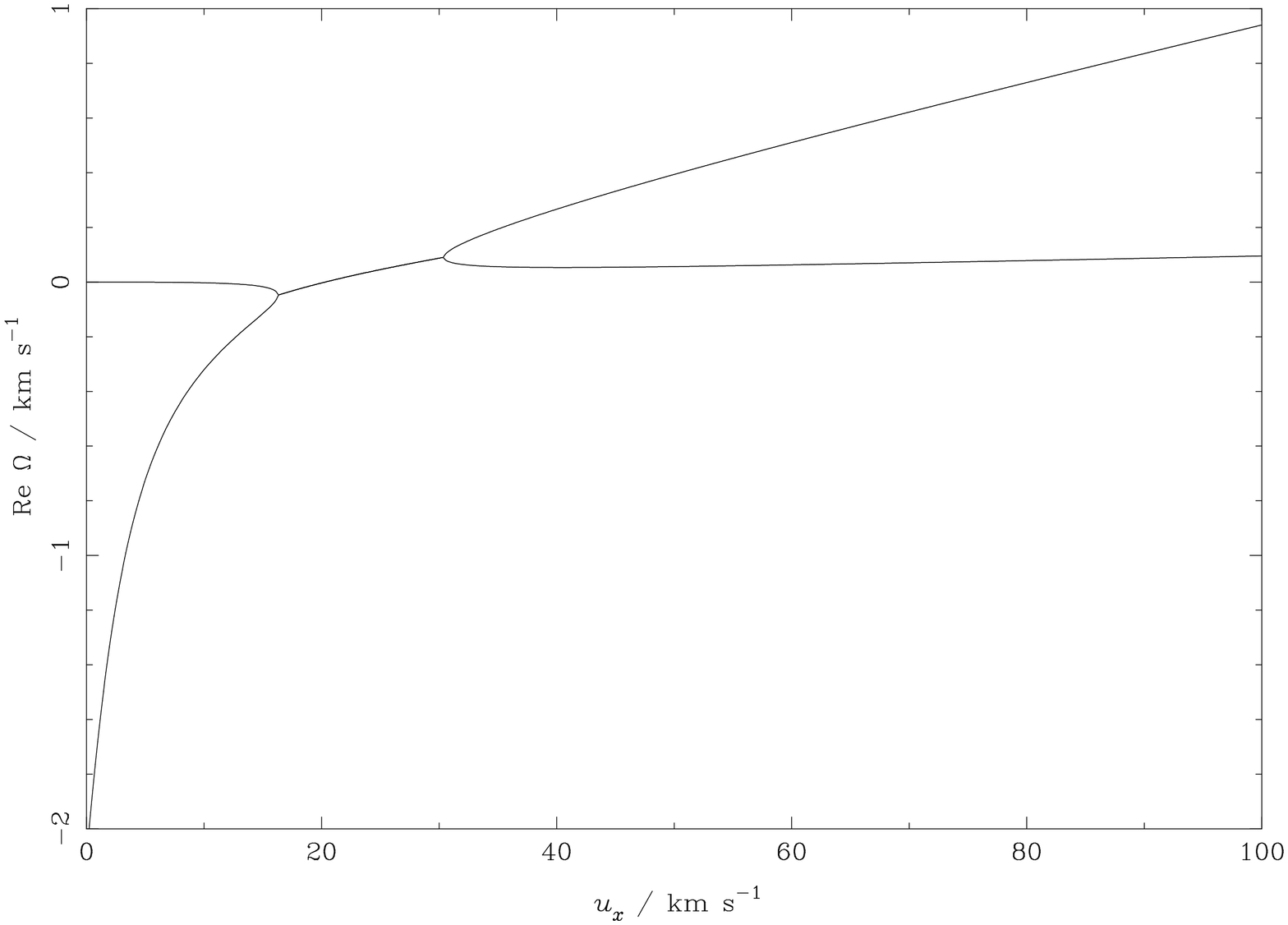}} &
\epsfysize=8cm\rotatebox{270}{\epsffile{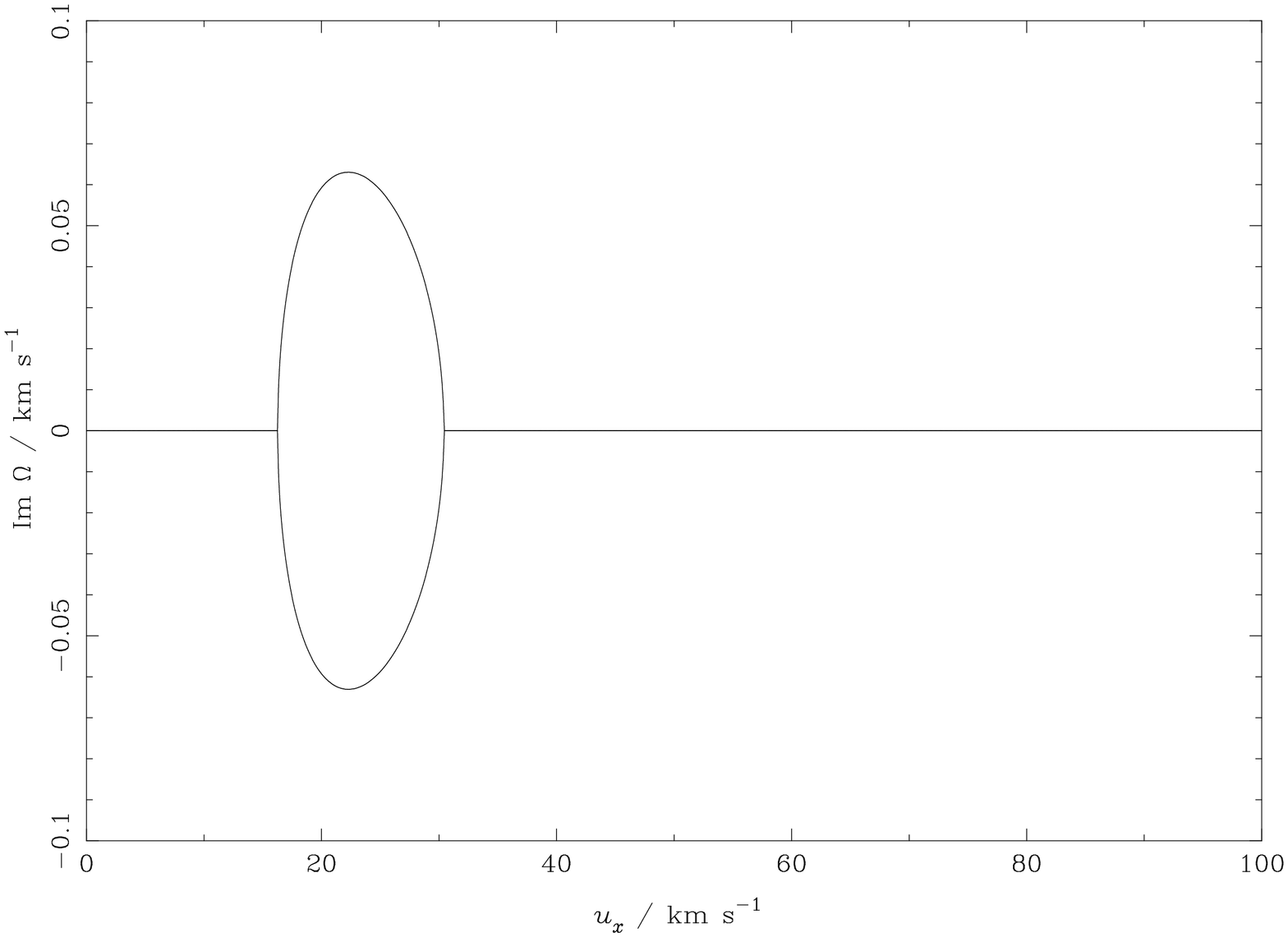}}
\end{tabular}
\end{centering}
\caption[]{Values of $\Omega$ for limit of long wavelengths, for
$v|A=1000{\rm\,km\,s^{-1}}$.  (a) and (b) show the real and imaginary
parts of two roots which are complex for part of the velocity range.
The remaining three roots are stable and not shown here.  The roots do
not tend to their asymptotic behaviour in this plot, as the maximal
slip velocity $u|{max}=100$ is small relative to the Alfv\'en velocity
(cf. the case of $v|A=20{\rm\,km\,s^{-1}}$ in
Figure~\protect\ref{f:long1}).  }
\label{f:long2}
\end{figure*}

\begin{figure*}
\begin{centering}
\begin{tabular}{ll}
(a) & (b) \\
\epsfysize=8cm\rotatebox{270}{\epsffile{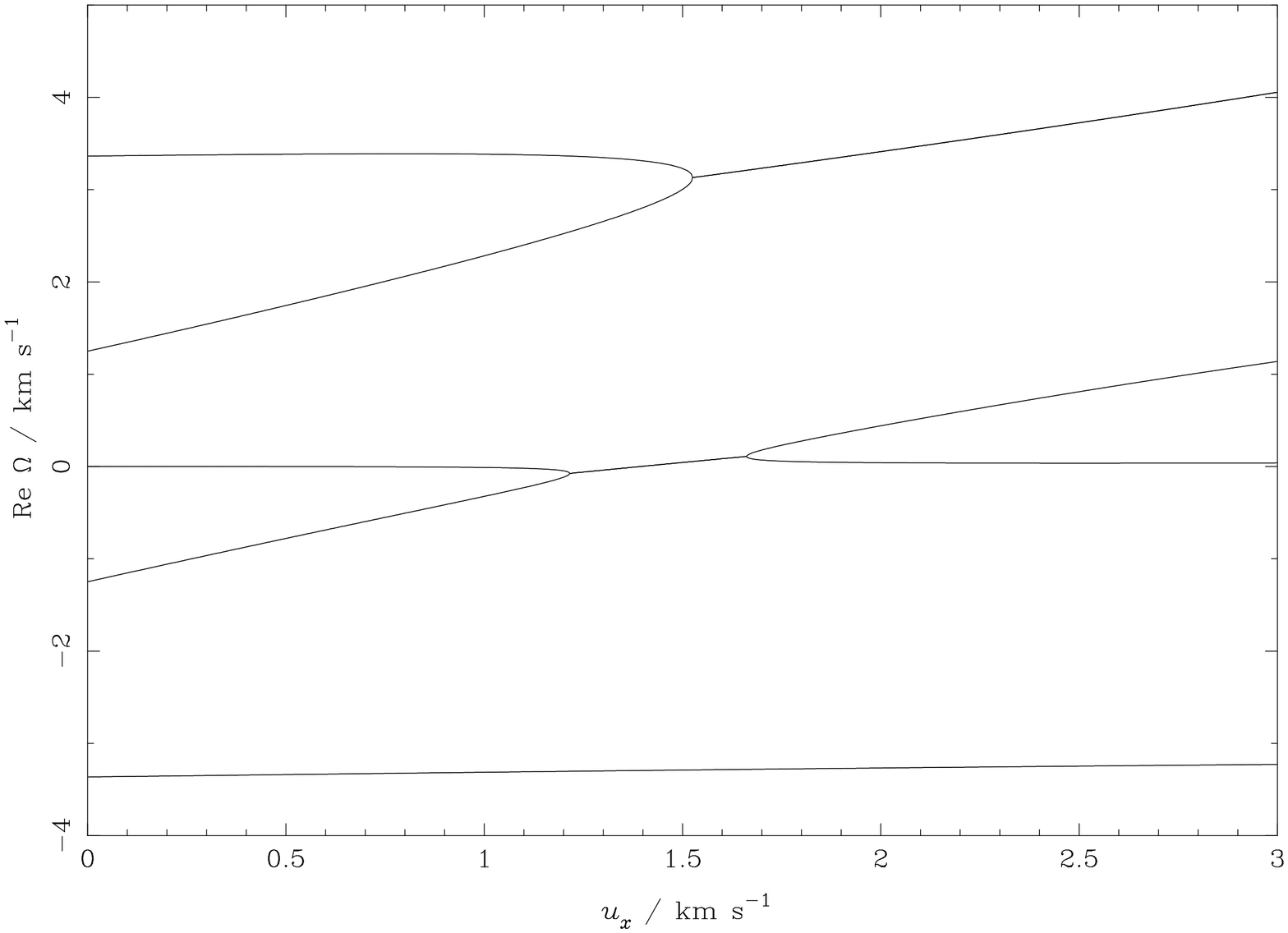}} &
\epsfysize=8cm\rotatebox{270}{\epsffile{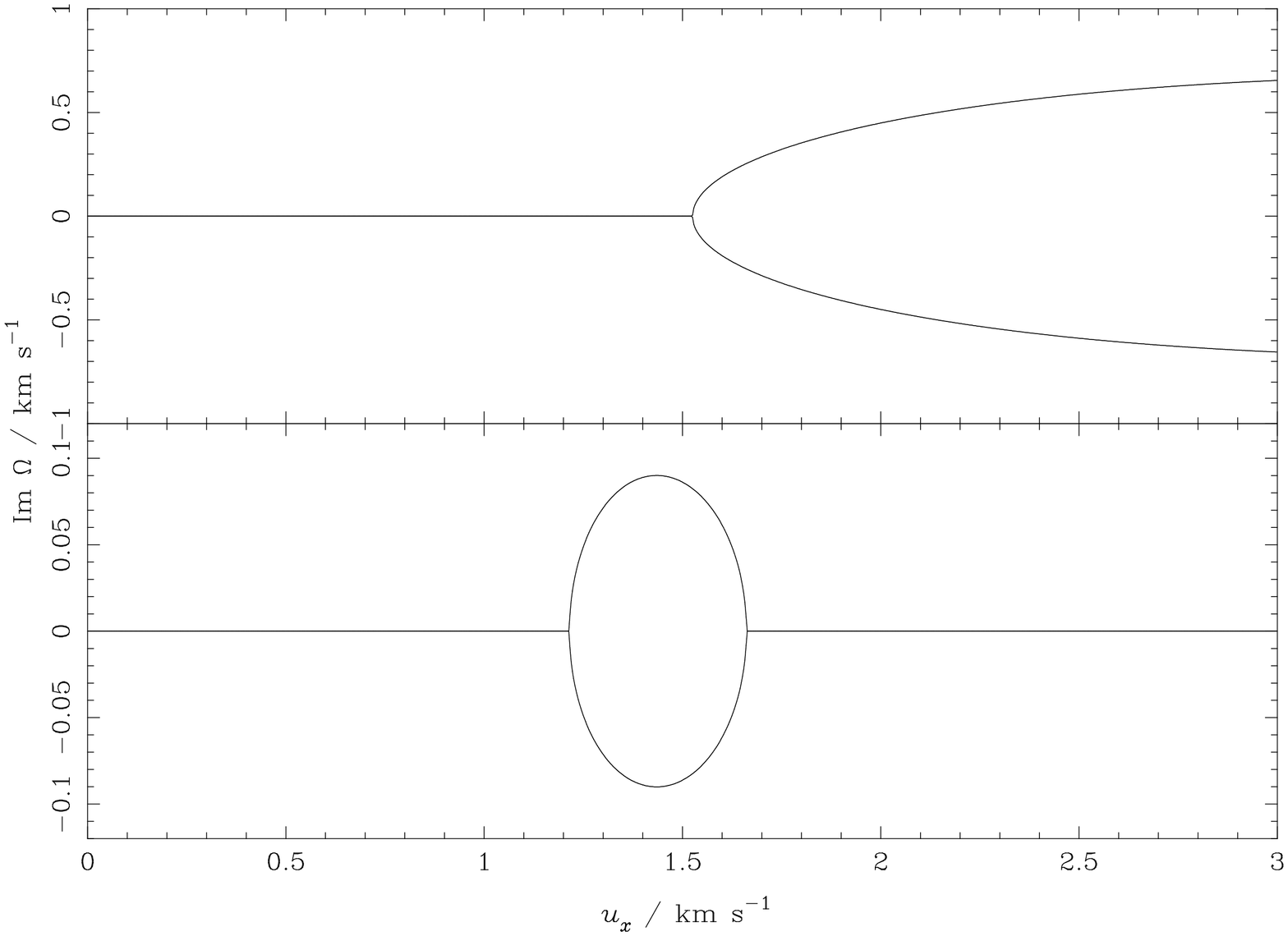}} \\
(c) & (d) \\
\epsfysize=8cm\rotatebox{270}{\epsffile{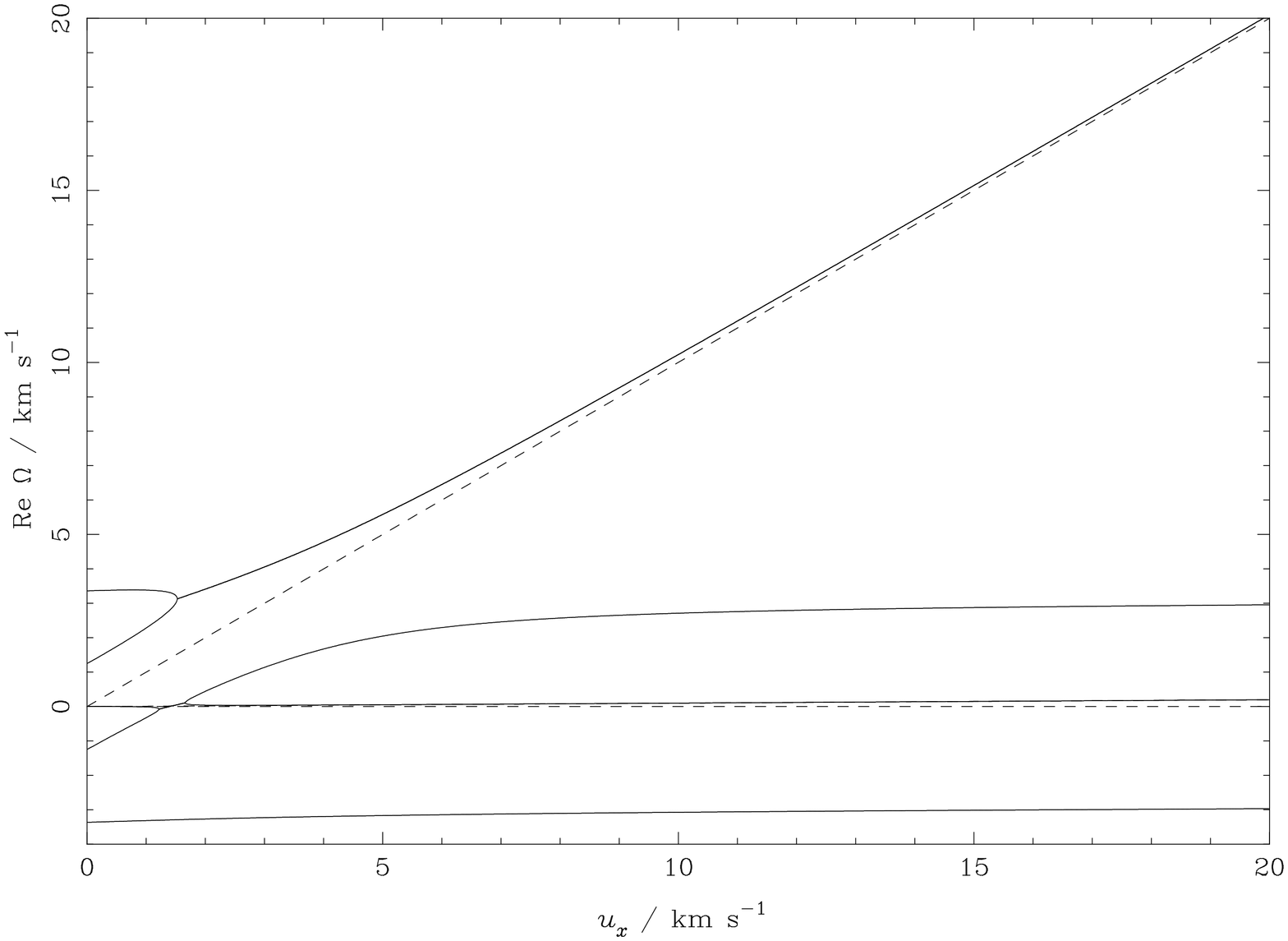}} &
\epsfysize=8cm\rotatebox{270}{\epsffile{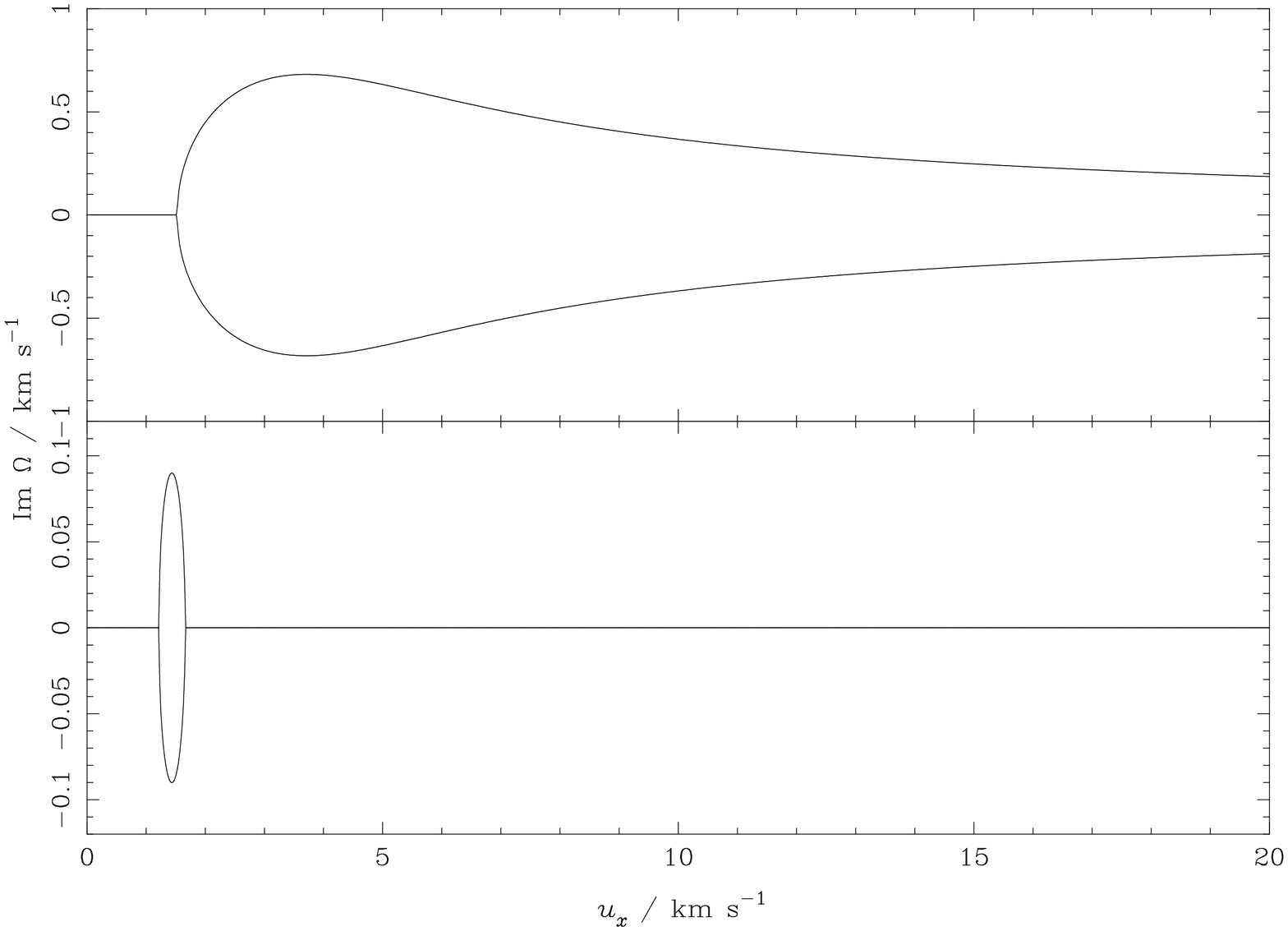}} \\
(e)\\
\epsfysize=8cm\rotatebox{270}{\epsffile{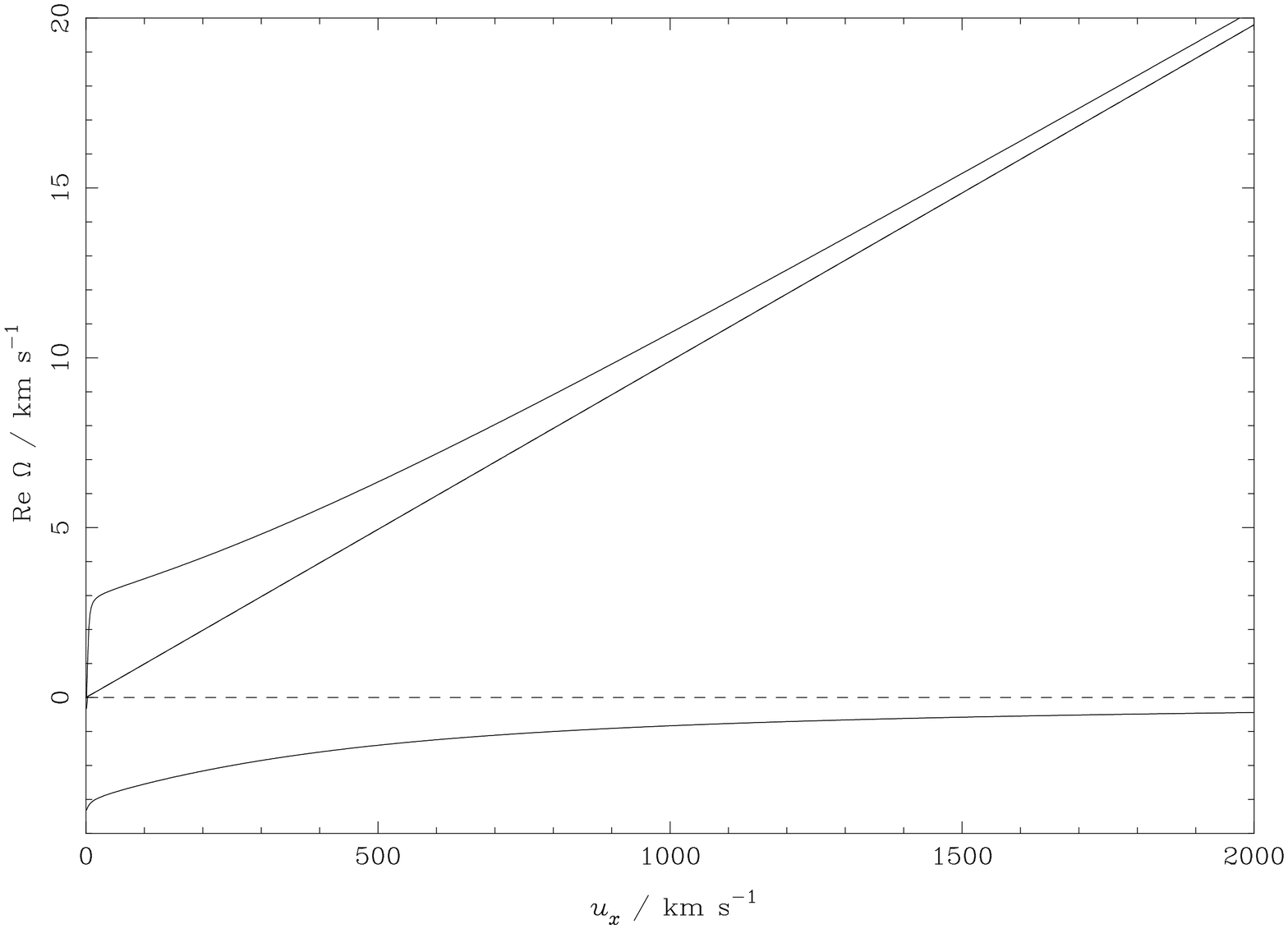}}
\end{tabular}
\end{centering}
\caption[]{Values of $\Omega$ in the long wavelength limit, for
$v|A=20{\rm\, km\,s^{-1}}$. (a) shows the real parts of the roots for
small $u_x$.  The regions in (a) where the real parts are distinct
correspond to stable roots with zero imaginary parts: (b) shows the
imaginary parts of the complex conjugate pairs.  (c) and (d) are the
same as (a) and (b), except that they are plotted for a rather wider
range of $u_x$, and the asymptotic behaviours of the roots are shown
as dashed curves.  At large $u_x$, the roots with the largest real
parts follow the expected asymptotic behaviour, but have finite
imaginary parts.  The next two roots should converge to a common
asymptotic limit, $\Omega\simeq 0.01u_x$, but while the variation of
the smaller of this pair is almost indistinguishable from this form
(indeed, it overlies the dashed curve showing the limit on the graph),
the larger lies well away from its asymptotic value, and convergence
is only clear at far larger $u_x$, as shown in (e). }
\label{f:long1}
\end{figure*}

As mentioned in the introduction, the current results are applicable
to mass loss from highly evolved stars, which are believed to be
driven by radiation pressure on dust particles
\cite{bowen88,macgs92,mmc96,simis01}.  The winds are subject to strong
perturbations as a result of stellar pulsation, and models have found
both radiation-pressure instabilities and others resulting from the
condensation of the dust grains, as well as radiative instabilities.
The structures which result from these instabilities may be important
in determining the properties of planetary nebula halos which will
form as the star ages further.  Mastrodemos \etal~\shortcite{mmc96}
find that slip between dust and gas components leads to formation of
dense shells at intermediate radii, but that these dissipate as the
wind moves away from the star.  Simis \etal~\shortcite{simis01},
however, find that dense shells survive to large radii, and that an
accurate treatment of dust-gas slip is essential in treatments of
late-type winds.

If, in the wind of a late-type star, ion-neutral friction is the main
source of coupling and the streaming speed is less than about
$30{\rm\,km\,s^{-1}}$, $\lambda$ can be calculated from the results
obtained by Osterbrock~\shortcite{oster} for the ion-neutral momentum
transfer cross section, so
\begin{equation}
\lambda \approx 1 \times 10^9 
\left( \frac{x}{10^{-6}} \right)
\left( \frac{\rho|n}{\rho|i} \right){\rm \,g^{-1}\,cm^3\,s^{-1}},
\end{equation}
where $x$ is the fractional ionization, i.e. the ratio of the number
density of ions to the number density of hydrogen nuclei. The value of
$k$ in the dimensionless units used in Figures~\ref{f:ex1}
and~\ref{f:max1} is related to the wavenumber in physical units
$k|{phys}$ by $k = k|{phys}/k_0$, where
\begin{equation}
k_0=\frac{\sqrt{2}\lambda\rho|i}{v|{Ai}}
\approx 3 \times 10^{-12}{\rm \, cm^{-1}}
\left( \frac{n|n}{10^{11} {\rm \, cm^{-3}}} \right)
\left( \frac{v|{Ai}}{10^3 {\rm\,km\,s^{-1}}} \right)^{-1}
\left( \frac{x}{10^{-6}} \right).
\end{equation}

To compare this with the scale-lengths of features in the winds, we
must have some idea of the appropriate values of $v|{Ai}$, $n|n$, $x$
and of the scale-lengths of interest.  SiO masers are found in the
outflows of some evolved stars at positions near those at which dust
is expected to form \cite[\eg{}]{kemball}.  The magnetic field
strength in the maser spots can be inferred with difficulty from the
observed polarization \cite[\eg{}]{watson}, although it is not certain
that the polarization is in fact caused by the Zeeman effect
\cite{wiebe}.  The resulting estimates of the magnetic field strengths
are in the range of 2 to 10 G \cite{elitzur,kemball}.  Pumping models
for these masers suggest that the neutral density in the spots is $n|n
\approx 10^{10}-10^{11} {\rm\,cm^{-3}}$ \cite{doel}.  If one percent
of the mass is contained in grains, $v|{Ai}$ could be as high as
$10^3{\rm\,km\,s^{-1}}$, but may be closer to $100{\rm\,km\,s^{-1}}$.
The size of individual maser spots is $\simeq 10^{12}$ cm
\cite[\eg{}]{kemball}, for which $k \approx 3 \times 10^{-13}
{\rm\,cm^{-1}}$ if the maser spot size corresponded to a half
wavelength of a perturbation. The fractional ionization is uncertain,
but it is unlikely that grains will form where carbon is ionized, and
some of the elements with low ionization potentials will be depleted
substantially from the gas phase due to the grain formation process.
Grain-neutral friction for a grain-neutral relative speed comparable
to the thermal speed of the neutral material is of similar magnitude
to the ion-neutral friction for $x$ between $10^{-7}$ and $10^{-6}$,
if the grains have a fractional abundance and size distribution
similar to those in the interstellar medium \cite[\eg{}]{baker}.

Therefore the SiO maser regions are probably large enough that their
overall properties correspond to our long wavelength limit.  To study
their stability in detail, we consider parameters characteristic of
the regions of late-type stellar winds with SiO maser spots, as
follows: $\rho|i/\rho|n = 0.01$, $c|n = 3{\rm\,km\,s^{-1}}$, $c|i =
0$, $v|A = 1000{\rm\,km\,s^{-1}}$ and $u =
0\mbox{--}100{\rm\,km\,s^{-1}}$.  We also assume that ${\bf u}$ is
perpendicular to magnetic field, i.e.\ ${\bf v}|A.{\bf u}=0$, and in
particular study the case where the wave vector, ${\bf k}$, bisects
the angle between ${\bf u}$ and ${\bf v}|A$.  Numerical solutions of
the long-wavelength dispersion relation are shown in
Fig.~\ref{f:long2}.  For comparison we also show in Fig.~\ref{f:long1}
the calculations for smaller Alfv\'en speed $v|A =
20{\rm\,km\,s^{-1}}$.  Note that for some slip speeds there are two
pairs of complex roots, but that for larger Alfv\'en velocities the
higher velocity roots are stable at all slip speeds.  We will refer to
the instability of the higher velocity roots, shown in
Figs~\ref{f:long1}a,b, as a type I instability, and that in
Figs~\ref{f:long2} and~\ref{f:long1}c,d as type II.

It will be noticed that as $v|A$ changes from 20 to
$1000{\rm\,km\,s^{-1}}$, the region of type II instability moves to
higher slip velocities.  For $v|A=20{\rm\,km\,s^{-1}}$, the maximum of
${\rm Im}(\Omega)$ is around $u_x \sim 1.5{\rm\,km\,s^{-1}}$ and for
$v|A=1000{\rm\,km\,s^{-1}}$ this maximum is around $u_x \sim 20-25
{\rm\,km\,s^{-1}}$.  The value of the maximum growth rate, however,
changes little, and is around 0.06--0.1$\rm\,km\,s^{-1}$.

For small Alfv\'en velocities (like 20 $\rm\,km\,s^{-1}$), a type I
instability is also present.  The regions of two instabilities overlap
to a degree, with the range of type II instability being for $u_x
\simeq 1.2\mbox{--}1.6{\rm\,km\,s^{-1}}$ and that for the type-I
instability being for $u_x \ga 1.5{\rm\,km\,s^{-1}}$.

The maser spots are typically at distances of 5--10 a.u.\@ from the
centres of the stars.  For an outflow speed of the order of
$10{\rm\,km\,s^{-1}}$, we conclude that an instability will have no
significant effect on the maser spots unless it grows on timescale of
a few years or less.  From Fig.~\ref{f:long1}b one can see that for
relatively small Alfv\'en velocities this is indeed the case, and such
maser spots would be destroyed by type I instability.  But as maser
spots exist at these radii, we can conclude that magnetic field (and
therefore the Alfv\'en velocity) cannot to be too small.  Indeed, for
Alfv\'en velocities around $1000 {\rm\,km\,s^{-1}}$, type-I
instability is suppressed and we only have to deal with type-II
instability (see Fig.~\ref{f:long2}).  This instability at its maximum
grows on timescale of order 5--10 years, but the range of streaming
velocities at which the instability has even this large a growth rate
is quite narrow, between 28 and $35{\rm\,km\,s^{-1}}$.  The streaming
speed of grains through a maser spot is uncertain and depends on the
fractional ionization and the grain size distribution; it may be that
in reality the streaming speed is too large or too small for the
instability to be driven at a rate high enough to destroy the maser
spot.  It will be important, as models of late-type stellar wind
including both the chemistry of grain formation and dynamics are
developed \cite[\eg{}]{macgs92,cherch,gail}, to determine the
relationship between the calculated grain-neutral streaming speed and
the input magnetic field.  These results are necessary for a full
application to stellar outflows of the present analysis.

\label{s:visc}
Finally, it is of interest to determine whether the short wavelength
analysis developed in previous sections can be applied to the
finer-scale properties of the maser spots.  This will be the case if
the widths of the resonant interactions are narrower than the spacing
of non-resonant modes, i.e. that
\begin{equation}
\frac{\lambda \rho|n}{k} \la V|{min},
\end{equation}
where $V|{min}$ is the smallest characteristic velocity of the problem
which should be $u \sim 1{\rm\,km\,s^{-1}}$, and we take $\rho|n$ in
the left hand side as it is larger than $\rho|i$.  This inequality
means that, in this example, short wavelengths are $2\pi/k \la
10^{10}{\rm\,cm}$.

This maximum wavelength must be compared to minimum wavelengths
required for the fluid approximations to be valid, and for the growth
of the two-fluid instability to outweigh wave damping as a result of
viscosity.  
%  For $T|e \le 7\times10^4{\rm\,K}$ the Coulomb
%logarithm may be approximated as
%\begin{equation}
%\log\Lambda\simeq 16-0.5\log \left(n|e\over 10^{-6}{\rm\,cm^{-3}}\right)
%+ 1.5 \log \left(T|e\over{\rm K}\right),
%\end{equation}
%with $n|e$ in units of ${\rm cm^{-3}}$ and $T|e$ in Kelvin,
%i.e. $\log\Lambda\simeq13.7$ for $n|e = 10^5{\rm\,cm^{-3}}$ and $T|e =
%10^3{\rm\,K}$, so that the electron-ion collision frequency is $200
%n|e T|e^{-1/2}$, and 
The maximum mean free paths for both electrons and ions are
\begin{eqnarray}
\lambda|{e,i} &\simeq& 10^5{\rm\,cm}\times\left(T\over 10^3{\rm\,K}\right)^2
\left(n|e\over 10^5{\rm\,cm^{-3}}\right)^{-1}
\end{eqnarray}
\cite{melrose}, while for neutral atomic species, the free path is
\begin{eqnarray}
\lambda|n &\simeq & {1\over n|n \sigma} \\
	&\simeq & 3\times10^4{\rm\,cm}\times
\left(n|n\over 10^{11}{\rm\,cm^{-3}}\right)^{-1},
\end{eqnarray}
assuming a geometric cross-section.  For longer wavelengths, the
finite collision frequencies will lead to viscous damping of waves in
a number of periods given by the ratio between the wavelength and the
free path, and any instability must grow rapidly enough to outweigh
this damping.  For a given slip velocity, the growth rates for the
instability we discuss are proportional to the density while the
viscous damping rates are proportional to the ratio of the square of
the frequency divided by the density, so damping will indeed only be
important for high-frequency waves in diffuse media.  In fact, at
these smaller scales further instability modes may also become
important \cite[driven by, e.g., the Hall effect,]{bt01}.

For the particular example under discussion, we require $10^5
{\rm\,cm} \ll \lambda|w \ll 10^{10} {\rm\,cm}$ which obviously can be
satisfied, so that our short wavelength analysis is applicable for a
substantial range of wavelengths.  In general, an appreciable range of
potentially unstable wavelengths remains for all temperatures $T \ll
10^6{\rm\,K}$.  While the shorter wavelengths have less direct
observational significance, their growth to nonlinear amplitudes will
likely have substantial effects on the overall flow structure.

\section{Discussion and conclusions}

\label{s:conclusion}
We have seen that the slip between different fluids in a multifluid
medium can drive instabilities.  At short wavelengths, several modes
of instability result from resonances between waves propagating in the
different fluids; these instabilities remain for longer wavelengths
but they become increasingly inter-coupled and less easy to
characterise.  In particular, neutral shear/ionized slow-mode
resonance instabilities will grow in almost all cases where there are
finite drift velocities.  The maximum growth rates we find are of
order the characteristic ion/neutral collision rates in gas without
internal slip, which are typically
\begin{equation}
\nu|{in} = 1.3\times10^{-10} n|n{\rm\,s^{-1}}
\end{equation}
\cite{oster,bz95}, rather than inversely proportional to density as is
the case for instabilities derived from leakage of magnetic flux as a
result of ambipolar diffusion.  This corresponds to a characteristic
lengthscale of $2.5\times10^{-4}
(v/1{\rm\,km\,s^{-1}})n|n^{-1}{\rm\,pc}$, where we note that the most
unstable wavelengths may be one hundredth of this
(Table~\ref{t:max1}).  The instabilities appear to be similar to the
two-stream instability of plasma flows \cite[\eg{}]{melrose,bingea00}.

In Section~\ref{s:wind}, we apply our analysis in its long-wavelength
limit to the overall properties of SiO maser spots in late-type
stellar winds.  We find that if the SiO spots are not to be subject to
violent instabilities, the slip velocity between the phases in these
regions must lie outside certain limits, or the magnetic field in
these spots must be very strong.

Our short wavelengths results also have important implications for
many other astrophysical systems.  In molecular clouds, we expect that
the ionized sound speed is small and that the magnetic field is
perpendicular to the inter-component slip velocity, but that this slip
velocity is not small.  This means that we always can find directions
such that inequality~(\ref{cond}) is satisfied.  It seems likely,
therefore, that molecular clouds are generically unstable to the
growth of slow-mode waves.  In the present analysis we can only
conjecture the non-linear endpoint of these instabilities, but the two
obvious possibilities -- fractionation of the phases as a `slugged'
flow, with consequent rapid loss of magnetic field support, or the
limiting of the wave spectrum at finite amplitude -- each have clear
practical and observational consequences for the ecology of the
interstellar medium.

Our present study is substantially simplified.  This has allowed us to
derive some rather general results.  In future work, we will model the
structure of the systems of interest more completely, including the
spatial structure of the background flow, more interacting phases
(e.g., treating electrons, ions, neutrals and a spectrum of sizes of
dust particle as independent species), variation of the frictional
constants with state, and the detailed nonlinear evolution of the
instabilities.

\section*{Acknowledgments}

PVT would like to thank the Royal Society and PPARC for support during
this work.  RJRW is supported by a PPARC Advanced Fellowship, and
thanks the Department of Physics and Astronomy in Leeds for
hospitality while this work was developed.  We wish to thank Tom
Hartquist for his role in suggesting this work, his continuing
interest in its development and useful comments on the manuscript.
Helpful comments from the anonymous referee and from J. Franco were
also much appreciated.

\appendix
\section{The Hermite-Biehler theorem}

\label{a:hb}
We quote here without proof the Hermite-Biehler theorem, a very
general result which underlies much of our analysis.  From theorem
$4'$ of Levin~\shortcite{lev64}, Chapter VII:
  
\newtheorem{hermite-biehler}{Theorem}
\begin{hermite-biehler}[Hermite-Biehler]
In order that the polynomial
\begin{equation}
w(z) = u(z)+iv(z),
\end{equation}
where $u(z)$ and $v(z)$ are real polynomials, does not have any roots
in the closed upper half-plane ${\rm Im}(z) \ge 0$, it is necessary
and sufficient that the following conditions are satisfied
\begin{enumerate}
\item the polynomials $u(z)$ and $v(z)$ have only simple real roots,
and these roots separate one another, \ie, between two successive
roots of one of these polynomials there lies exactly one root of the
other;
\item at some point $x_0$ of the real axis
\begin{equation}
v'(x_0)u(x_0)-v(x_0)u'(x_0) < 0,
\end{equation}
where $u'(z) = du(z)/dz$ and $v'(z) = dv(z)/dz$.
\end{enumerate}
\end{hermite-biehler}

\label{lastpage}\end{document}